\documentclass[manuscript]{acmart}

\usepackage{xcolor}  %
\usepackage{colortbl}
\usepackage{subcaption} 
\usepackage{pifont}    %
\usepackage{makecell}  %
\usepackage{utfsym}
\usepackage{multirow}
\usepackage{enumitem}
\usepackage{graphicx}
\usepackage{placeins} %
\usepackage{booktabs}  %
\usepackage{tabularx}  %
\usepackage{siunitx}   %
\definecolor{OliveGreen}{RGB}{0,100,0}

\AtBeginDocument{%
  }

\acmJournal{PACMHCI}
\acmVolume{10}
\acmNumber{XXX}
\acmDOI{}
\setcopyright{none}
\begin{document}
\sloppy 
\authorsaddresses{}
\title{PalmSpace: Towards a Versatile On-Palm Interaction Space through Unified Touch Modeling}

\author{Chentao Li}
\affiliation{%
  \department{Department of Automation}
  \institution{Department of Automation, Tsinghua University}
  \city{Beijing}
  \state{Beijing}
  \country{China}
}

\author{Mingze Gao}
\affiliation{%
  \department{Academy of Arts \& Design}
  \institution{Academy of Arts \& Design, Tsinghua University}
  \city{Beijing}
  \country{China}
}

\author{Runze Sun}
\email{sunrz22@mails.tsinghua.edu.cn}
\affiliation{%
  \institution{Department of Automation, Tsinghua University}
  \city{Beijing}
  \country{China}
}

\author{Zhaoguo Wang}
\affiliation{%
  \department{Department of Automation, BNRist}
  \institution{Department of Automation, Tsinghua University}
  \city{Beijing}
  \country{China}
}

\author{Jianjiang Feng}
\authornote{Corresponding author.}
\affiliation{%
  \department{Department of Automation}
  \institution{Department of Automation, Tsinghua University}
  \city{Beijing}
  \country{China}
}

\author{Jie Zhou}
\affiliation{%
  \department{Department of Automation, BNRist}
  \institution{Department of Automation, Tsinghua University}
  \city{Beijing}
  \country{China}
}

\renewcommand{\shortauthors}{Li et al.}

\begin{abstract}

As smart glasses and lightweight MR devices become increasingly practical, input remains a key challenge. 
The bare palm is an always-available, tactile, and proprioceptively accessible surface, but it has neither an explicit coordinate system nor embedded touch sensing. 
Prior on-palm systems typically expose isolated touch events, discrete regions, continuous trajectories, or task-specific gestures, limiting the palm's ability to support precise selection and gesture manipulation through a common input representation.
We present PalmSpace, a wrist-worn infrared system that exposes mode-aware, body-referenced absolute input on the bare palm without per-user sensing calibration. 
At the interaction level, PalmSpace jointly represents contact occurrence, interaction mode, and palm-referenced absolute location; at the model level, it learns these coupled outputs through a shared real-time representation.
In leave-one-participant-out evaluation with 17 participants, PalmSpace achieved 6.7 mm mean localization error, 98.9\% contact detection accuracy, and 96.7\% F1 for four-class interaction-state recognition. 
User studies further demonstrated absolute pointing and dragging, eyes-free digit input, and representative multi-finger controls including scrolling and pinch-based map manipulation. 
These results show that a morphologically variable bare palm can function as a transferable, mode-aware interaction surface.
\end{abstract}


\begin{teaserfigure}
  \centering
  \includegraphics[width=0.95\textwidth]{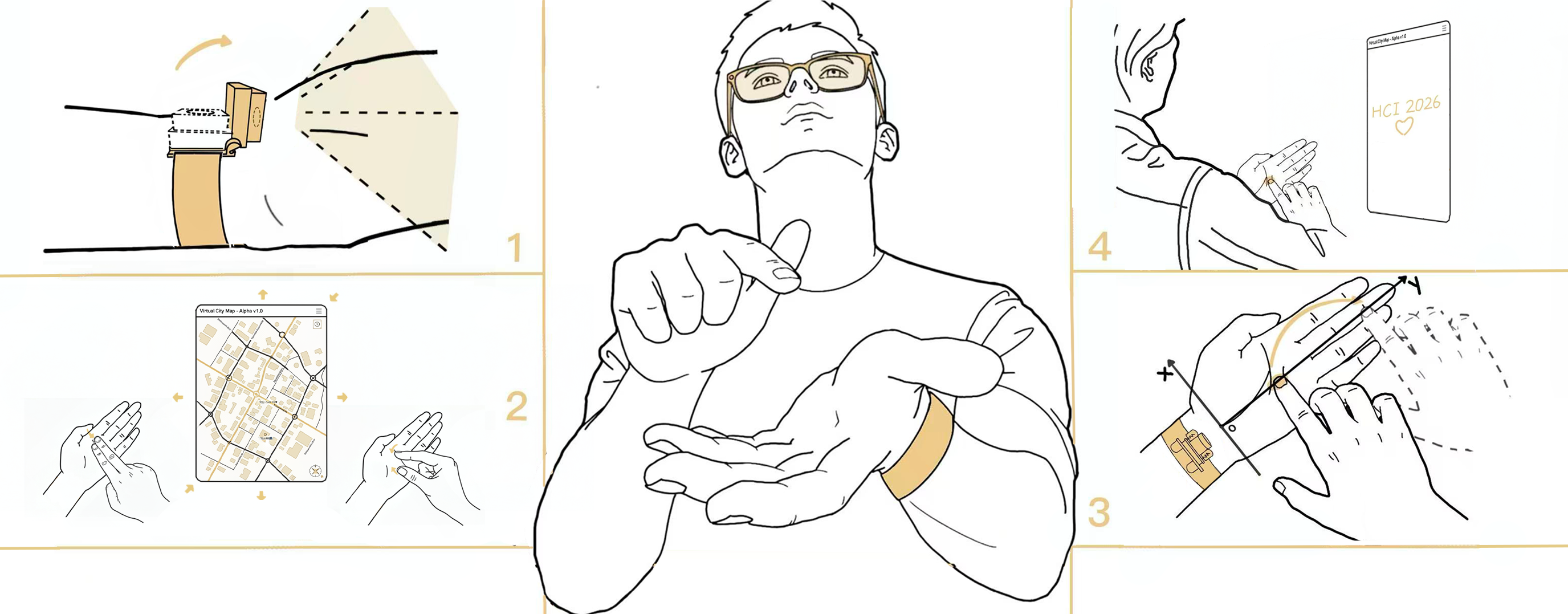}
  \caption{Diagram of the core concept of PalmSpace. 
  PalmSpace turns the palm into a structured and eyes-free interaction space.
  The four panels demonstrate: (1) a flip-up camera design, 
  (2) a structured interaction space that supports complex map manipulation requiring both single-finger localization and multi-finger scroll/pinch input;
  (3) absolute positioning across the entire palm area including fingers, and 
  (4) support for precision-demanding interactions like handwriting.}
  \label{fig:teaser}
  \Description{Four panels show the PalmSpace wristband with its camera flipped up, map manipulation using single- and multi-finger palm input, tracked touch locations across the palm and fingers, and a handwriting example drawn on the palm.}
\end{teaserfigure}

\maketitle

\section{Introduction}

Head-worn displays are again drawing attention as smart glasses and lightweight mixed-reality devices move closer to practical everyday use \cite{lee2018interaction,he2025palmpad}. Yet input remains a central challenge. Everyday wearable interaction demands techniques that are always available, socially unobtrusive, visually undemanding, and precise enough for both quick commands and fine-grained manipulation. While mice and touchpads remain the gold standard for precise pointing, they depend on dedicated hardware and stable external surfaces, making them poorly suited to lightweight head-worn use in mobile settings. Existing alternatives such as eye tracking, mid-air hand input, and compact wearable controllers avoid bulky external hardware, but often come with trade-offs such as visual demand, fatigue, lack of haptic feedback, or reliance on relative rather than absolute control \cite{lee2024gazepointar,mayer2020enhancing,qin2023selecting,kim2023worldpoint,nickel2003pointing,shen2024mousering,kubo2022ring,streli2024touchinsight}. This makes on-body input particularly attractive for smart glasses: it keeps the interaction surface always available on the body itself while preserving tactile and proprioceptive cues.

Among on-body input surfaces, the palm is especially promising: it is always available, naturally private, and rich in tactile and proprioceptive cues. 
Yet unlike a touchscreen, the bare palm has neither an explicit coordinate system nor embedded sensing for determining when and how input occurs. 
In terms of practicality, camera-based approaches are inherently viewpoint-dependent, making touch inference vulnerable to limited fields of view, self-occlusion, oblique viewing angles, and weak depth cues \cite{mollyn2024egotouch,he2025palmpad}. Systems augmented with projectors or depth sensors can improve sensing quality, but their additional bulky hardware substantially reduces deployment practicality \cite{harrison2011omnitouch,dezfuli2012palmrc}. Even wearable solutions often still rely on user-specific calibration or training, while offering limited robustness in touch detection and localization \cite{stearns2018touchcam,wang2015palmgesture,zhang2016skintrack,wang2015palmtype}.
More fundamentally, prior systems tend to expose only one layer of input semantics: discrete touch zones, isolated touch events, continuous positions, or task-specific gestures \cite{wang2015palmtype,stearns2018touchcam}. Most also lack explicit modeling of multi-finger modes alongside absolute position.
As a result, the palm is still treated more as a collection of sparse events than as a structured interaction space.
What is missing is a common representation of palm interaction that jointly encodes three tightly coupled properties: \textit{whether} contact occurs, \textit{where} it occurs in body-referenced absolute coordinates, and \textit{what} interaction mode it belongs to. 
For example, in smart-glasses map interaction, users may first use touch detection and absolute localization to select or drag toward a region of interest, and then switch to multi-finger input for zooming and rotation. 
Representing these properties together allows precise targeting and gesture control to coexist on the same bare-palm surface.

To this end, we present \textit{PalmSpace}, a wrist-worn system that provides unified on-palm input at two complementary levels. At the interaction level, it exposes contact semantics, interaction mode, and palm-referenced absolute location through a common representation. At the model level, it learns interaction state and location through a shared multi-task representation rather than independent task-specific pipelines.
We instantiate this formulation with a wrist-worn infrared sensing setup, a palm-normalized coordinate space that transfers across users without per-user sensing calibration, and a real-time multi-task Transformer. Together, these components support precise single-finger input and representative multi-finger gesture control on the same bare-palm surface.

In offline evaluation using leave-one-participant-out cross-validation, PalmSpace achieved a mean absolute localization error of 6.7 mm, 98.9\% accuracy for binary contact detection, and a 96.7\% F1-score for four-class interaction-state recognition. An additional ablation on unseen outdoor data shows that the unified model improves both localization and touch recognition over independently trained task-specific models while requiring only one deployed model.

To evaluate PalmSpace in realistic settings, we conducted user studies of Fitts's Law pointing, eyes-free digit entry, handwriting, list navigation, and map manipulation, together with outdoor evaluations under daytime and low-light conditions.

In summary, the main contributions of this paper are as follows:

\begin{itemize}[leftmargin=*]
\item We present \textbf{PalmSpace}, an on-palm input system that jointly represents contact occurrence, interaction mode, and body-referenced absolute location, turning the bare palm into a mode-aware interaction surface without per-user sensing calibration.

\item We instantiate this representation through wrist-worn infrared sensing, palm-normalized coordinates, and a real-time multi-task Transformer. Compared with independently trained task-specific models, unified learning improves zero-shot outdoor localization and touch recognition while avoiding a second deployed model.

\item Through cross-user evaluation and indoor/outdoor user studies, we characterize PalmSpace across absolute pointing, eyes-free symbolic input, and representative multi-finger gesture control, achieving 6.7\,mm localization error, 98.9\% binary contact accuracy, and 96.7\% four-class F1.
\end{itemize}

\section{Related Work}

\begin{table*}[h]
  \centering
  \caption{Comparison of sensing setup, input representation, and evaluation scope across existing on-palm techniques. PalmSpace combines contact-aware continuous absolute positioning with multi-finger mode recognition without per-user sensing calibration. (`C': continuous, `D': discrete)}
  \renewcommand{\arraystretch}{1.4}
  \resizebox{\textwidth}{!}{%
  \begin{tabular}{lcccccccccccc}
  \toprule
  \textbf{Study} & \textbf{Year} & \textbf{\makecell{Sensor\\Type}} & \textbf{\makecell{Sensor\\Placement}} & \textbf{\makecell{Real-world\\Evaluation}} & \textbf{\makecell{Sensing\\Calibration Free}} & \textbf{\makecell{Eyes\\Free}} & \textbf{\makecell{Absolute\\Positioning}} & \textbf{\makecell{2D Finger\\Tracking}} & \textbf{\makecell{Multi-Finger\\Mode}} & \textbf{Error} & \textbf{\makecell{Touch\\Detection}} & \textbf{\makecell{Touch/Gesture\\Accuracy}} \\
  \midrule
  Kohli and Whitton \cite{kohli2005haptic} & 2005 & Magnetic trackers & Finger+Back & \textcolor{red}{\usym{2715}} & \textcolor{red}{\usym{2715}} & \textcolor{red}{\usym{2715}} & \textcolor{OliveGreen}{\ding{51}} & \textcolor{OliveGreen}{C} & \textcolor{red}{\usym{2715}} & -- & \textcolor{OliveGreen}{\ding{51}} & -- \\
  PalmRC \cite{dezfuli2012palmrc} & 2012 & Depth camera & Back Shoulder & \textcolor{red}{\usym{2715}} & \textcolor{OliveGreen}{\ding{51}} & \textcolor{OliveGreen}{\ding{51}} & \textcolor{red}{\usym{2715}} & \textcolor{red}{D} & \textcolor{red}{\usym{2715}} & 28 mm & \textcolor{OliveGreen}{\ding{51}} & $>90\%$ \\
  PalmGesture \cite{wang2015palmgesture} & 2015 & IR camera+laser & Wrist & \textcolor{red}{\usym{2715}} & \textcolor{red}{\usym{2715}} & \textcolor{OliveGreen}{\ding{51}} & \textcolor{OliveGreen}{\ding{51}} & \textcolor{OliveGreen}{C} & \textcolor{red}{\usym{2715}} & -- & \textcolor{red}{\usym{2715}} & $>90\%$ \\
  PalmType \cite{wang2015palmtype} & 2015 & IR sensors & Wrist & \textcolor{red}{\usym{2715}} & \textcolor{OliveGreen}{\ding{51}} & \textcolor{OliveGreen}{\ding{51}} & \textcolor{red}{\usym{2715}} & \textcolor{red}{D} & \textcolor{red}{\usym{2715}} & -- & \textcolor{OliveGreen}{\ding{51}} & 74.5\% \\
  SkinTrack \cite{zhang2016skintrack} & 2016 & Electrodes & Wrist+Finger & \textcolor{red}{\usym{2715}} & \textcolor{red}{\usym{2715}} & \textcolor{OliveGreen}{\ding{51}} & \textcolor{OliveGreen}{\ding{51}} & \textcolor{OliveGreen}{C} & \textcolor{red}{\usym{2715}} & 24.1 mm & \textcolor{OliveGreen}{\ding{51}} & 92.6\% \\
  TouchCam \cite{stearns2018touchcam} & 2018 & IR+camera+IMU & Finger & \textcolor{red}{\usym{2715}} & \textcolor{red}{\usym{2715}} & \textcolor{OliveGreen}{\ding{51}} & \textcolor{red}{\usym{2715}} & \textcolor{red}{D} & \textcolor{red}{\usym{2715}} & -- & \textcolor{OliveGreen}{\ding{51}} & $\sim96\%$ \\
  EgoTouch \cite{mollyn2024egotouch} & 2024 & RGB camera & Headset & \textcolor{OliveGreen}{\ding{51}} & \textcolor{OliveGreen}{\ding{51}} & \textcolor{red}{\usym{2715}} & \textcolor{red}{\usym{2715}} & \textcolor{OliveGreen}{C} & \textcolor{red}{\usym{2715}} & -- & \textcolor{OliveGreen}{\ding{51}} & 95.6\% \\
  Palmpad \cite{he2025palmpad} & 2025 & RGB camera & Headset & \textcolor{red}{\usym{2715}} & \textcolor{OliveGreen}{\ding{51}} & \textcolor{red}{\usym{2715}} & \textcolor{red}{\usym{2715}} & \textcolor{OliveGreen}{C} & \textcolor{red}{\usym{2715}} & -- & \textcolor{OliveGreen}{\ding{51}} & 97.0\% \\
  \rowcolor{gray!20}
  PalmSpace (Ours) & 2026 & IR camera & Wrist & \textcolor{OliveGreen}{\ding{51}} & \textcolor{OliveGreen}{\ding{51}} & \textcolor{OliveGreen}{\ding{51}} & \textcolor{OliveGreen}{\ding{51}} & \textcolor{OliveGreen}{C} & \textcolor{OliveGreen}{\ding{51}} & 6.7 mm & \textcolor{OliveGreen}{\ding{51}} & 98.9\%/96.7\% \\
  \bottomrule
  \end{tabular}}
  \label{tab:related_work}
\end{table*}

\subsection{Input Methods for Head-Worn Interaction}

Head-worn interaction requires input techniques that are efficient, easy to use, private, socially acceptable, and suitable for eyes-free use in daily activities~\cite{xu2025fingerglass,yoon2016wearable}. At the same time, they must support both diverse commands and sufficiently precise control.
Gaze-based interaction remains a dominant modality, using AR/MR headsets~\cite{lee2024gazepointar,wang2024g,yan2024voila} or smartphone cameras~\cite{mayer2020enhancing,qin2023selecting} for interaction. However, it suffers from the ``Midas Touch'' problem~\cite{jacob1991use}, ocular fatigue~\cite{lee2024gazepointar}, and lower precision than traditional mice~\cite{majaranta2014eye}. Robust gaze input also often depends on dedicated eye tracking or rich sensing configurations, which are difficult to support in lightweight head-worn settings~\cite{lee2024gazepointar,wang2024g,yan2024voila}.
Mid-air hand interaction offers an intuitive alternative, including pointing-direction input~\cite{kim2023worldpoint,fang2024watchthis,hu2010hand,vogel2005distant,li2026egopoint} and fingertip estimation~\cite{romaniak2020nimble,malik2004visual}. Yet, these methods depend on line of sight and provide little haptic feedback, complicating fine motor control and increasing fatigue during prolonged use~\cite{cheng2022comfortable}.
Physical devices provide tangible feedback, but come with different trade-offs. Pen-based inputs offer high precision, but are less portable because they require an additional handheld device and often rely on relative input with mobile devices or external tracking systems~\cite{luthi2022deltapen,matulic2020pensight,hwang2013magpen,romat2021flashpen,maierhofer2024tiptrack}. Smart rings improve portability~\cite{kubo2022ring,kienzle2014lightring,kienzle2021electroring,oh2017anywheretouch,shen2024mousering}, yet often sacrifice absolute positioning and wearing comfort. Smartphones~\cite{babic2018pocket6,debarba2012lop,pietroszek2014smartcasting,pietroszek2015tiltcasting,he2023trackpose,li2024swiveltouch,yu2025touchforce} and smartwatches~\cite{katsuragawa2016watchpoint} similarly support portable interaction, with different capabilities and constraints in sensing and control.

\subsection{On-Hand Input Methods}

Prior research has explored a range of techniques for turning the hand or palm into an input surface. 
For smart-glasses, however, two challenges remain central: whether palm input can be sensed with a lightweight and wearable setup, and whether the palm can support versatile, structured interaction.

\textit{Practical sensing methods.}
Early systems demonstrated the feasibility of palm interaction by augmenting the environment or the body with additional sensing hardware. 
Kohli and Whitton \cite{kohli2005haptic} used magnetic tracking to monitor the index finger on the opposite palm, but the method was prone to drift and required careful calibration. 
OmniTouch \cite{harrison2011omnitouch}, WatchSense \cite{sridhar2017watchsense}, and LumiWatch \cite{xiao2018lumiwatch} combined depth sensing and projection to support on-body touch interaction. 
While these systems showed the promise of interactive skin, they depended on depth cameras, projectors, or bulky mounted hardware, which increase cost, size, and deployment burden \cite{mollyn2024egotouch}. 
Other wearable sensing approaches reduced some of this hardware overhead. SkinTrack \cite{zhang2016skintrack} used RF-based triangulation for continuous 2D tracking, but required user calibration and was sensitive to skin conditions. 
TouchCam \cite{stearns2018touchcam} achieved high touch recognition accuracy by instrumenting the finger, while HandPad \cite{lu2024handpad} relied on attachments on both wrist and fingers. 
Although effective, these solutions introduce instrumentation on the fingers or body that reduces practicality. 
More recently, head-mounted MR systems such as EgoTouch \cite{mollyn2024egotouch} and Palmpad \cite{he2025palmpad} have shown that finger-palm contact can be detected using egocentric cameras.
However, such approaches remain inherently viewpoint-dependent: touch inference is unstable under self-occlusion, oblique viewing angles, or when the hand leaves the camera view,
as evidenced by Palmpad’s reliance on the Quest 2’s multi-camera API for localization.
Taken together, prior work has explored multiple sensing routes, yet practical palm interaction for smartglass use remains constrained by bulky hardware, finger instrumentation, calibration requirements, or viewpoint dependency.

\textit{From touch events to absolute palm interaction.}
Beyond sensing, a second challenge is how the palm is represented as an input space. 
Several systems have leveraged the palm as a proprioceptively accessible surface for eyes-free interaction \cite{bergstrom2019human}. 
PalmRC \cite{dezfuli2012palmrc} and PalmType \cite{wang2015palmtype}, for example, showed that the palm can support remote control and text entry without visual attention. 
However, PalmType relied on discrete infrared proximity sensors and was limited to detecting predefined key regions rather than continuous coordinates. 
PalmGesture \cite{wang2015palmgesture} moved toward continuous input using an infrared camera and laser line projector, but required users to define the interaction area by touching four corners and imposed strict posture constraints on finger orientation. 
These limitations illustrate that many prior systems still treat the palm either as a set of discrete touch zones or as a calibrated surface requiring explicit setup, rather than as a robust absolute interaction space. 
For head-worn use, however, practical palm interaction requires more than knowing that a touch occurred; it also requires reliable absolute localization on the palm surface without per-user calibration.

\textit{From single-finger input to expressive multi-finger interaction.}
A further limitation of existing on-skin interfaces lies in interaction expressiveness. 
Most compact wearable palm-input systems primarily support single-finger tapping or pointing \cite{wang2015palmtype,stearns2018touchcam,zhang2016skintrack,xiao2018lumiwatch}. 
By contrast, studies have shown that users naturally transfer multi-finger gesture metaphors, such as pinching and two-finger sliding, from smartphones to skin surfaces \cite{weigel2014more}. 
Systems that support simultaneous multi-point input, such as iSkin \cite{weigel2015iskin} and SkinMarks \cite{weigel2017skinmarks}, typically require stretchable overlays or electronic tattoos attached directly to the skin, which reduce tactile sensitivity and complicate everyday deployment. Projection-based on-skin interaction systems can support multi-point input, but generally require bulky sensing hardware \cite{harrison2011omnitouch}.
In the realm of compact, wrist-worn devices (e.g., SkinTrack \cite{zhang2016skintrack}, LumiWatch \cite{xiao2018lumiwatch}), signal ambiguity and occlusion have largely restricted interactions to single-finger pointing or discrete tapping.
As a result, prior work has largely left compact palm interfaces at the level of sparse input events, without explicitly modeling richer interaction modes in a unified way. 
In particular, existing systems rarely address together three essential dimensions of palm interaction: whether contact occurs, where it occurs in absolute palm coordinates, and which interaction mode it belongs to, such as single-finger versus multi-finger input.

\textit{Summary and Research Gap.}
Table~\ref{tab:related_work} summarizes how PalmSpace compares with prior on-palm interaction systems in terms of sensing setup, calibration requirements, localization capability, interaction expressiveness, and evaluation scope. 
Taken together, prior work has improved palm input sensing, but still falls short of making the palm a practical and general interaction space. Existing systems typically address only part of the problem---such as touch detection, localization, or a specific gesture family---while often relying on bulky hardware, per-user calibration, or sustained visual attention. As a result, palm input remains fragmented into isolated events and task-specific pipelines rather than functioning as a unified interaction space. PalmSpace addresses this gap by jointly modeling \emph{contact existence}, \emph{absolute on-palm location}, and \emph{interaction mode} in a lightweight, calibration-free wrist-worn system.

\section{PalmSpace Overview}

\subsection{Proof-of-Concept Hardware} \label{sec:hardware}

PalmSpace is a wrist-mounted infrared vision system for sensing bimanual finger-to-palm interaction. 
It is designed to capture fine-grained palm contact cues and support three tasks: touch detection, touch localization, and gesture recognition. 
At this stage, we implement PalmSpace as a proof-of-concept prototype to validate sensing feasibility and system design under stable imaging conditions.

\textbf{Why a wrist viewpoint?}
A wrist-worn camera provides a direct and stable view of the hand and palm during interaction, making it well suited for contact sensing. 
Compared with head-mounted viewpoints~\cite{mollyn2024egotouch,streli2023structured} or external tracking APIs~\cite{dupre2024tripad}, the wrist perspective more directly captures the contact region and is therefore better suited for subtle touch sensing~\cite{xi2026wristpp,wang2026pianoband}. 
More importantly, unlike most prior wrist-mounted vision systems, which mainly focus on single-hand pose estimation or finger tracking~\cite{hu2020fingertrak,xi2026wristpp,kim2012digits,li2025finetype}, PalmSpace targets bimanual finger-to-palm interaction, a fundamentally different task that requires sensing touch events, touch locations, and interaction gestures rather than only posture. 

\begin{figure}[ht]
  \begin{subfigure}[c]{0.46\textwidth}
    \centering
    \includegraphics[width=\linewidth]{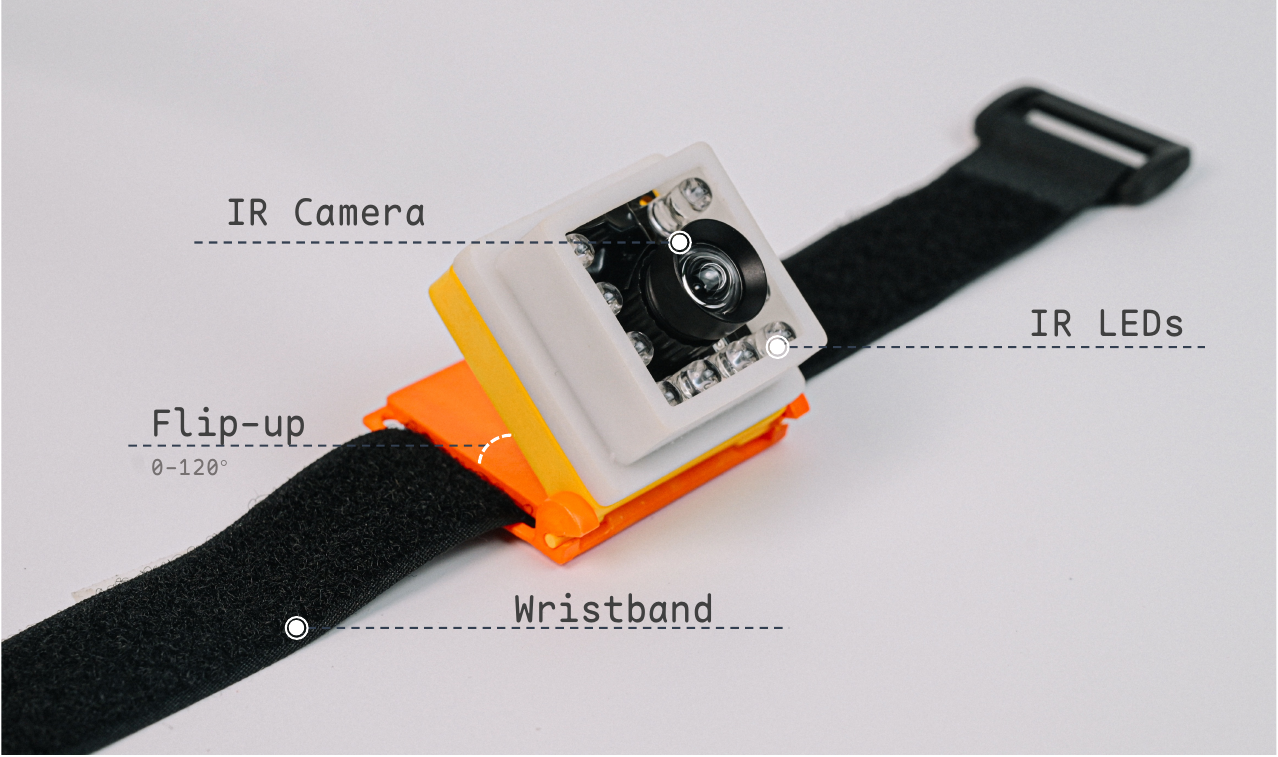}
    \caption{Hardware prototype}
    \label{fig:hardware}
\end{subfigure}
\hfill 
\begin{subfigure}[c]{0.46\textwidth}
    \centering
    \includegraphics[width=\linewidth]{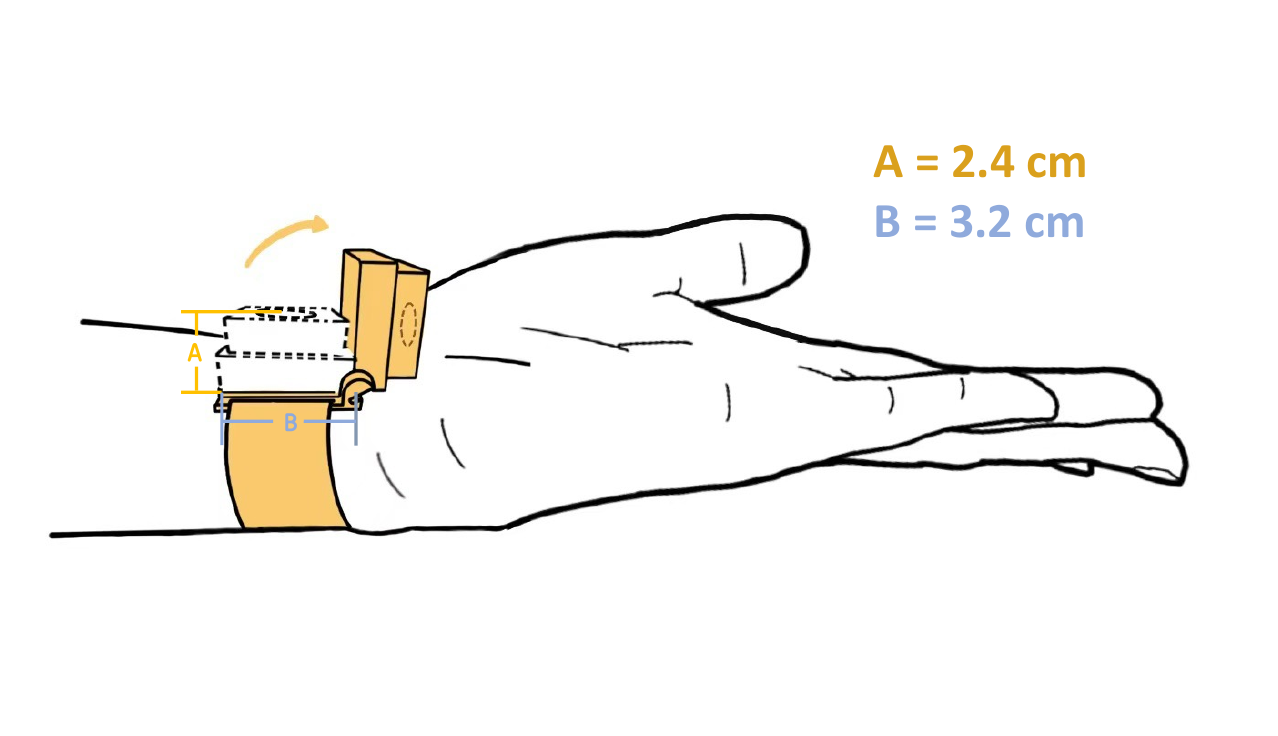}
    \caption{Wristband's form factor}
    \label{fig:wristband_form}
\end{subfigure}
\caption{PalmSpace's hardware prototype overview.}
\Description{Two views of the PalmSpace prototype show the infrared camera and LED hardware, followed by the wristband worn on the inner wrist with the camera module raised above the palm.}
\end{figure}

\textbf{Prototype design.}
Our wristband (Fig.~\ref{fig:hardware}) integrates an infrared camera (OV2710) with a 3.5~mm wide-angle lens (100$^\circ$ FOV), capturing 1920$\times$1080 infrared images at 30~FPS. 
Ten 850~nm infrared LEDs are embedded in a 3D-printed casing to provide active illumination, enabling robust infrared imaging that reduces ambient light interference and reveals clear palm details. 
To reduce self-occlusion during wrist bending, we introduce a 120$^\circ$ mechanical flip-up structure, similar to the design of imoo Z7~\cite{imoo_z7}, allowing the camera to maintain visibility across a wide range of hand poses. 
The device is worn on the dorsal wrist in standby mode and rotated to the inner wrist during sensing. 
The prototype measures 2.4~cm in height and 3.2~cm in width from the side, and weighs 22~g excluding cables. 
At this stage, PalmSpace is implemented as a wired prototype for reliable image capture and synchronization; future wireless integration is feasible given commercial wearable camera systems~\cite{tonsen2020high,tapwithus,Wristcam_2020}.

\subsection{Versatile Interaction Modes} \label{sec:intermode}

\begin{figure*}[ht]
  \includegraphics[width=\textwidth]{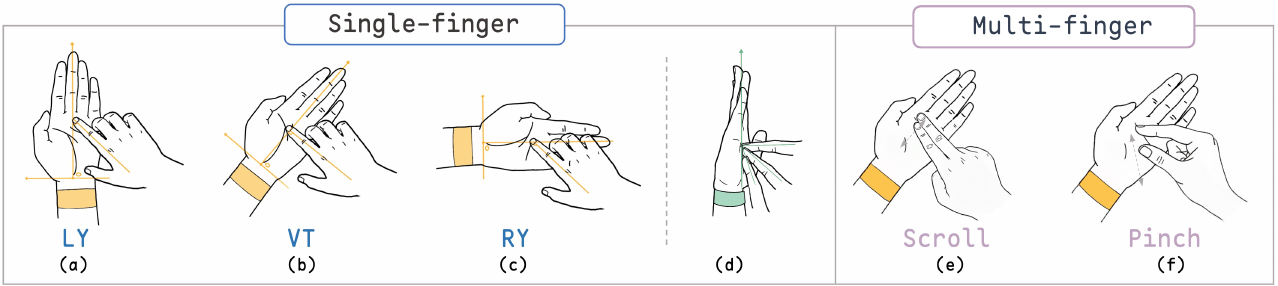}
  \caption{Illustration of finger-palm interaction modes. (a)-(c) Single-finger modes: users operating in LY, VT, and RY modes respectively. (d) The pitch angle between the finger and palm surface can be freely adjusted. (e)-(f) Multi-finger gestures: Scroll and Pinch are performed on the palm surface.}
  \label{fig:interaction_mode}
  \setcounter{subfigure}{0}
  \refstepcounter{subfigure}\label{subfig:mode_a}
  \refstepcounter{subfigure}\label{subfig:mode_b}
  \refstepcounter{subfigure}\label{subfig:mode_c}
  \refstepcounter{subfigure}\label{subfig:pitch_d}
  \refstepcounter{subfigure}\label{subfig:scroll_e}
  \refstepcounter{subfigure}\label{subfig:pinch_f}
  \Description{Six panels illustrate three single-finger yaw orientations, free adjustment of finger pitch, a two-finger scrolling gesture, and a pinch gesture performed on the palm.}
\end{figure*}

To realize the versatile interaction space introduced earlier, PalmSpace structures the palm as a unified interaction space rather than a collection of isolated touch events. We organize this space along two dimensions: finger configuration (single-finger vs. multi-finger) and contact surface (palm vs. finger surface), enabling both precise input and expressive manipulation. 
For single-finger input, we define three natural posture variants—Left Yaw (LY), Vertical Touch (VT), and Right Yaw (RY)—based on the finger’s yaw angle relative to the palm (Fig.~\ref{subfig:mode_a}-\ref{subfig:mode_c}). These variants capture common on-body touch orientations while allowing pitch to vary freely (Fig.~\ref{subfig:pitch_d}), and can occur on both the palm and finger surfaces. 
For multi-finger input, PalmSpace supports two palm-based modes, Scroll and Pinch (Fig.~\ref{subfig:scroll_e} and~\ref{subfig:pitch_d}), whose larger and smoother contact area better supports stable multi-finger control. Together, these modes turn the palm into a structured interaction space for symbolic input, precise spatial control, and compound manipulation.

\section{PalmSpace Method} \label{sec:algorithm}

PalmSpace is built on a simple premise: applications should receive palm input through a common representation rather than disconnected touch, localization, and gesture pipelines.
PalmSpace therefore unifies input at two levels. At the interaction level, it couples discrete state semantics with continuous absolute location on the palm. At the model level, it learns both outputs through a shared multi-task representation from wrist-worn images.

\subsection{Unified Touch Modeling Framework}
\label{sec:framework}

The interaction representation integrates three tightly coupled properties: 
(1) \textit{contact-state classification}, which distinguishes non-contact from active touch; 
(2) \textit{interaction-mode recognition}, which identifies whether the current interaction corresponds to single-finger input or a supported multi-finger mode; and 
(3) \textit{touch localization}, which estimates the absolute interaction position in a palm-normalized coordinate space. 
Instead of deploying independent task-specific models, PalmSpace learns these properties within one end-to-end model, enabling real-time interaction without per-user sensing calibration.
This unified formulation structures the palm as an interaction space with hierarchical interaction awareness. 
Single-finger input supports precise pointing and target selection, while multi-finger input supports richer manipulations such as scrolling and pinch-based control. 
In our implementation, the model predicts four interaction states: \textit{Non-contact}, \textit{Single-finger}, \textit{Scroll}, and \textit{Pinch}. 
These four states jointly encode both contact presence and interaction mode. 
At the same time, the model regresses the 2D location of the primary interaction point, thereby coupling discrete interaction semantics with continuous spatial input.

\subsection{Palm-Normalized Touch Representation}
\label{sec:normalized}

A central challenge of touch localization on the palm is the large anatomical variation across users. 
Directly regressing fingertip positions in absolute image or physical coordinates makes predictions highly sensitive to hand size differences. 
As a result, even the same semantic touch location on the palm (e.g., the tip of the middle finger) can correspond to substantially different coordinates across users. 
In our dataset, the gap between the smallest and largest hands reaches 4.5~cm, causing significant bias for direct coordinate regression.
To address this issue, PalmSpace represents touch locations in a palm-relative normalized coordinate system. 
The horizontal coordinate is centered at the wrist midline and normalized by palm breadth, while the vertical coordinate is referenced to the palm base and normalized by palm length. 
This maps the palm into a standardized space with lateral range $[-0.5,0.5]$ and longitudinal range $[0,1]$, shifting localization from user-specific absolute prediction to proportional position estimation within the palm. 
This representation improves cross-user consistency, supports calibration-free learning, and enables the learned interaction space to transfer naturally across devices with different screen sizes.

\subsection{Multi-Task Transformer Architecture}
\label{sec:architecture}

\begin{figure*}[t]
  \centering
  \includegraphics[width=0.94\linewidth]{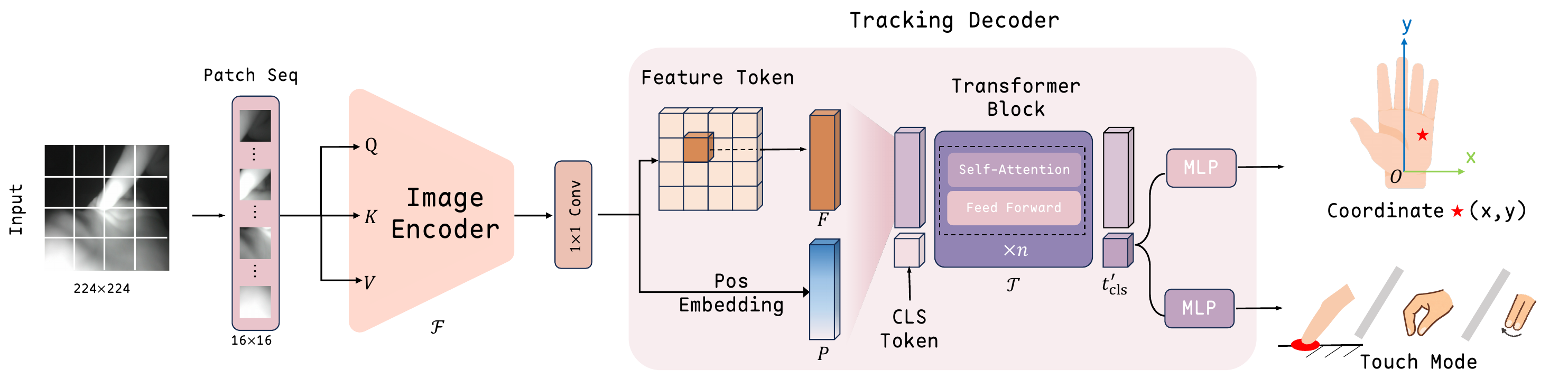}
  \caption{PalmSpace network architecture.}
  \label{fig:network}
  \Description{A multi-task Transformer pipeline processes wrist-camera palm images, extracts shared features, and jointly predicts contact and gesture class together with a normalized two-dimensional interaction location.}
\end{figure*}

To instantiate the unified formulation above, we design a multi-task Transformer that jointly predicts interaction state and touch location from wrist-worn images (Fig.~\ref{fig:network}). 
The model contains three components: a visual feature extractor, a lightweight Transformer interaction module, and two task-specific prediction heads.

Given an input image, we first extract visual features using a pre-trained ViT-B/16 backbone~\cite{dosovitskiy2020image}. 
The backbone output is reshaped into a spatial feature map and projected by a $1\times1$ convolution to obtain
$F \in \mathbb{R}^{d_{\text{model}}\times H \times W}$, where $d_{\text{model}}=256$. 
To preserve spatial information for touch localization, we add absolute 2D sinusoidal positional embeddings $P$ to the projected tokens.
The resulting tokens are then processed by a lightweight Transformer module $\mathcal{T}$ with three encoder layers. 
A learnable class token $t_{\text{cls}}$ is prepended to aggregate a representation of the interaction state:
\begin{equation}
F' = \mathcal{T}(F + P),
\end{equation}
yielding the output sequence
\begin{equation}
[t'_{\text{cls}}, f'_1, f'_2, \ldots, f'_{H\times W}].
\end{equation}

We use the transformed class token $t'_{\text{cls}}$ as the shared latent representation and attach two parallel heads. 
The classification head predicts four interaction states: \textit{Non-contact}, \textit{Single-finger}, \textit{Scroll}, and \textit{Pinch}. 
The regression head predicts the normalized 2D coordinates $(x,y)$ of the interaction point.

For single-finger input, the regression target is the active contact location. For multi-finger gestures, PalmSpace regresses the index finger as the primary interaction point. 
This proxy-based design is sufficient for the two gesture types studied here: in Scroll, index-finger displacement captures gesture direction and offset; in Pinch, the index finger provides a stable reference relative to the palm center for estimating radial and angular change.

\subsection{Training Objective}
\label{sec:loss}

PalmSpace is trained end-to-end with a multi-task objective that jointly optimizes interaction-state classification and touch localization:
\begin{equation}
\mathcal{L}_{\text{total}} = \lambda_{\text{cls}} \mathcal{L}_{\text{cls}} + \lambda_{\text{reg}} \mathcal{L}_{\text{reg}},
\end{equation}
where $\mathcal{L}_{\text{cls}}$ and $\mathcal{L}_{\text{reg}}$ denote the classification and regression losses, respectively.
For interaction-state prediction, the model classifies each sample into one of four states: \textit{Non-contact}, \textit{Single-finger}, \textit{Scroll}, and \textit{Pinch}. We use the standard cross-entropy loss:
\begin{equation}
\mathcal{L}_{\text{cls}} = -\frac{1}{N}\sum_{i=1}^{N}\sum_{c=1}^{K}\mathbb{1}(y_i=c)\log(\hat{p}_{i,c}),
\end{equation}
where $N$ is the batch size, $K=4$ is the number of interaction states, $y_i$ is the ground-truth label, and $\hat{p}_{i,c}$ is the predicted probability for class $c$.
For touch localization, we regress the interaction point using a masked mean squared error:
\begin{equation}
\mathcal{L}_{\text{reg}} = \frac{1}{\sum_{i=1}^{N} M_i}\sum_{i=1}^{N} M_i \cdot \|\tilde{C}_i - C_i\|_2^2,
\end{equation}
where $\tilde{C}_i$ and $C_i$ are the predicted and ground-truth coordinates, and $M_i=\mathbb{1}(y_i \neq \text{Non-contact})$ masks out non-contact samples, which have no valid spatial target.
We set $\lambda_{\text{cls}}=1$ and $\lambda_{\text{reg}}=100$ to account for the smaller numerical scale of normalized coordinate regression and to maintain effective joint optimization. Additional implementation details are provided in Appendix~\ref{app:system_implementation}.

\section{Offline Evaluation}

This section evaluates PalmSpace for joint touch mode recognition and touch localization, covering the dataset and validation protocol, detailed results across contact regions and interaction types, and ablation studies of key design components.
\subsection{Data Collection}
\subsubsection{Pipeline.}
To evaluate PalmSpace, we collected a comprehensive dataset of finger-to-palm interactions covering the variations shown in Fig.~\ref{fig:interaction_mode}. 
Each sample was annotated with four labels: (1) touch state (touch vs.\ non-touch), (2) gesture category (Single, Scroll, or Pinch), (3) interaction coordinates on the hand surface, and (4) palm size for coordinate normalization. 
Our setup used a wrist-mounted IR camera for model input and an overhead mirrorless camera for ground-truth generation. 
To recover physical interaction coordinates, we defined a planar calibration area around the hand with AprilTag markers \cite{apriltag} and mapped the overhead view to the physical coordinate system via homography estimation. 
Fingertip keypoints were tracked in the overhead view using MediaPipe \cite{zhang2020mediapipe} and transformed into metric palm-plane coordinates as localization ground truth. 
This markerless design avoids attaching sensors or reflective markers, reducing visual artifacts while preserving realistic bare-hand interaction. 
Details are provided in Appendix~\ref{app:datacollect}.

\subsubsection{Participants and Procedure.}
We recruited 17 right-handed volunteers (11 male, 6 female; aged 20--34 years, $M=24.4$, $SD=3.3$). 
To ensure anthropometric diversity, we recorded hand dimensions: hand length ranged from 155 to 200 mm ($M=181.1$, $SD=12.0$), and hand breadth ranged from 65 to 85 mm ($M=77.8$, $SD=5.2$), consistent with prior reports \cite{khan2017determination}. 
The study included 9 sessions: 6 single-finger sessions (2 areas $\times$ 3 modes), 2 multi-finger sessions (Scroll and Pinch), and 1 non-contact session. 
Single-finger data were collected on both the palm and finger regions, whereas multi-finger gestures were collected on the palm only. 
Participants performed gestures naturally without strict pose constraints to better reflect realistic touchpad-like interactions. 
The non-contact session included hovering and random hand motions as negative samples for touch detection. 
After removing samples with ground-truth tracking failures, the final dataset contained 63,903 images from 17 participants, including 35,727 single-finger samples, 11,645 multi-finger samples, and 16,531 non-contact samples. 
Each participant completed the procedure in about 30 minutes.

\subsection{Validation Setup}

We adopt a leave-one-participant-out (LOPO) cross-validation strategy. 
Given $N=17$ participants, the dataset is partitioned into $N$ folds. In each fold, data from $N-1$ users are used for training, while the remaining user serves as the unseen test set. 
This subject-independent protocol rigorously evaluates the model's generalization ability to new users.

We evaluate model performance from three perspectives:
\begin{itemize}[leftmargin=*]
    \item \textbf{Contact Region:} performance on single-touch detection and localization over the \textbf{Palm} and \textbf{Fingers}.
    \item \textbf{Single-finger Touch Mode:} performance under the three finger orientation patterns (\textbf{LY}, \textbf{VT}, and \textbf{RY}) defined in Sec.~\ref{sec:intermode}.
    \item \textbf{Multi-finger Gesture:} performance on dynamic multi-finger interactions, specifically \textbf{Scroll} and \textbf{Pinch}.
\end{itemize}

For touch gesture recognition, we report the F1-score. 
For absolute localization, we use the Mean Absolute Error (MAE) and Standard Deviation (SD) on the $x$/$y$ axes ($x_{\text{error}}$, $y_{\text{error}}$), as well as the Euclidean distance error ($l_{\text{error}}$). 
Localization is evaluated only on samples with correctly detected valid touch events (Single, Scroll, or Pinch). 
And the metrics are micro-averaged from all test folds.

\subsection{Performance}
\begin{figure*}[htbp]
  \centering
  \begin{subfigure}[t]{0.18\textwidth}
    \centering
    \includegraphics[height=2.45cm]{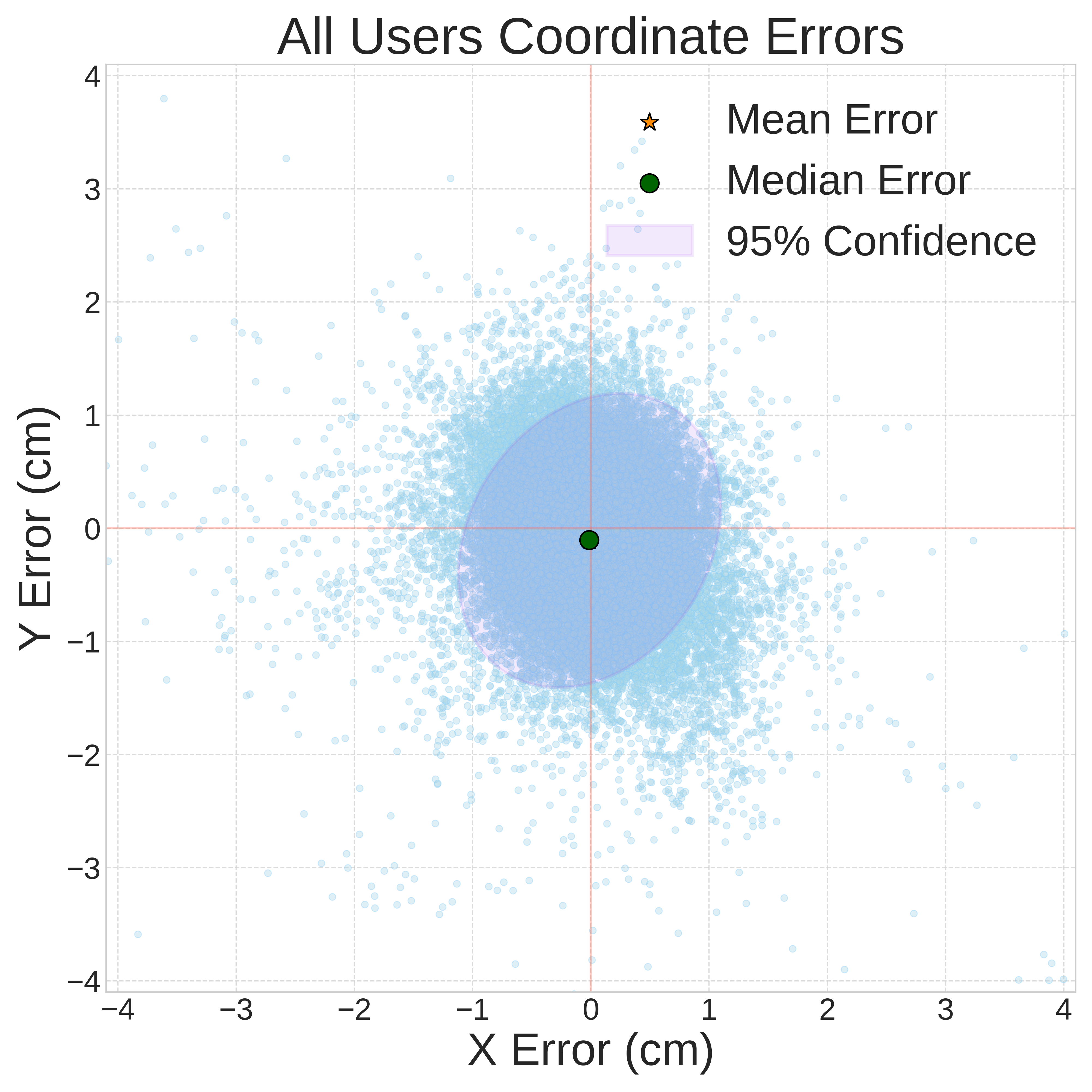}
    \caption{Prediction errors in the $x$--$y$ plane.}
    \label{fig:error_1}
  \end{subfigure}
  \hfill
  \begin{subfigure}[t]{0.22\textwidth}
    \centering
    \includegraphics[height=2.45cm]{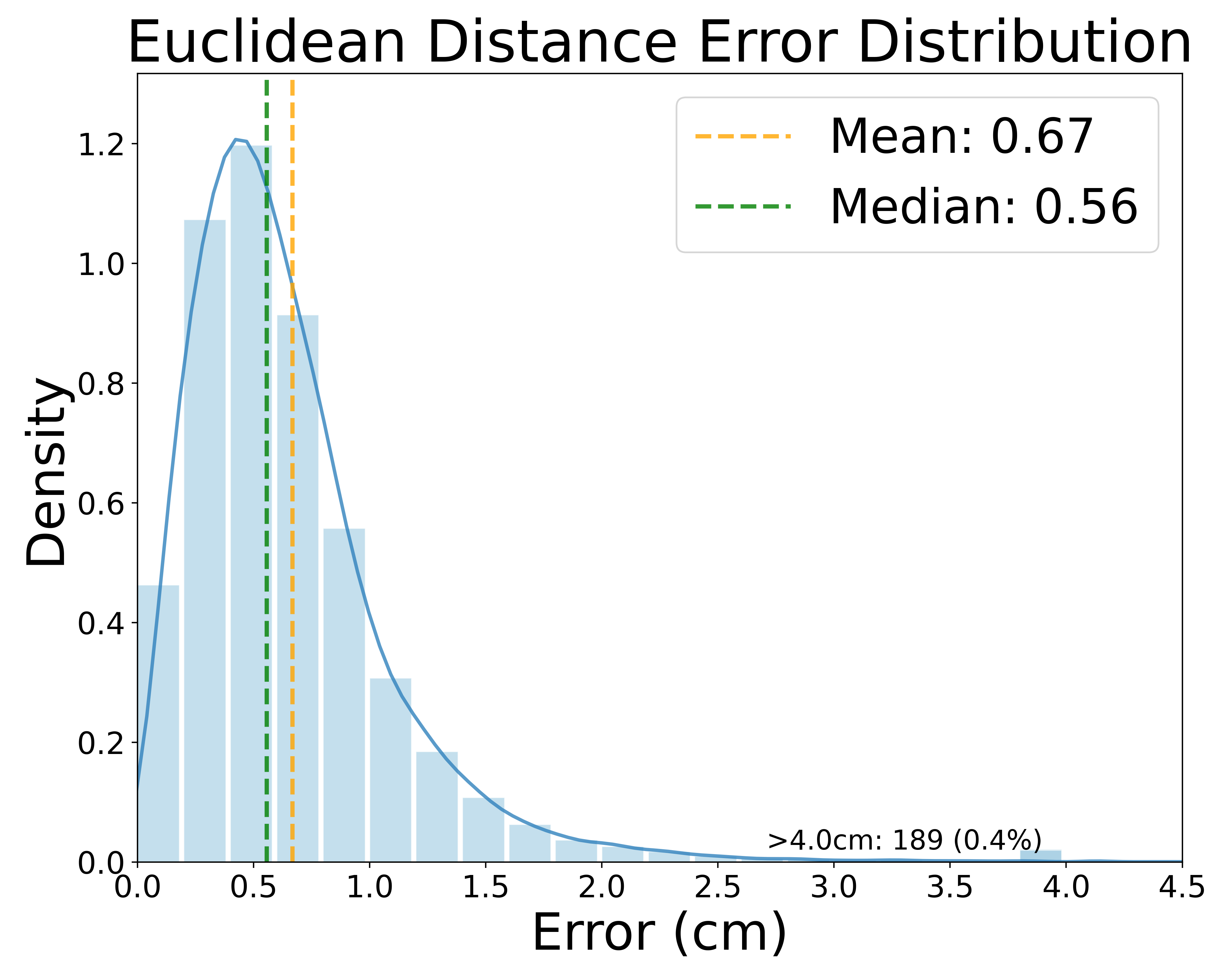}
    \caption{Distribution of Euclidean localization error.}
    \label{fig:error_2}
  \end{subfigure}
  \hfill
  \begin{subfigure}[t]{0.27\textwidth}
    \centering
    \includegraphics[height=2.45cm]{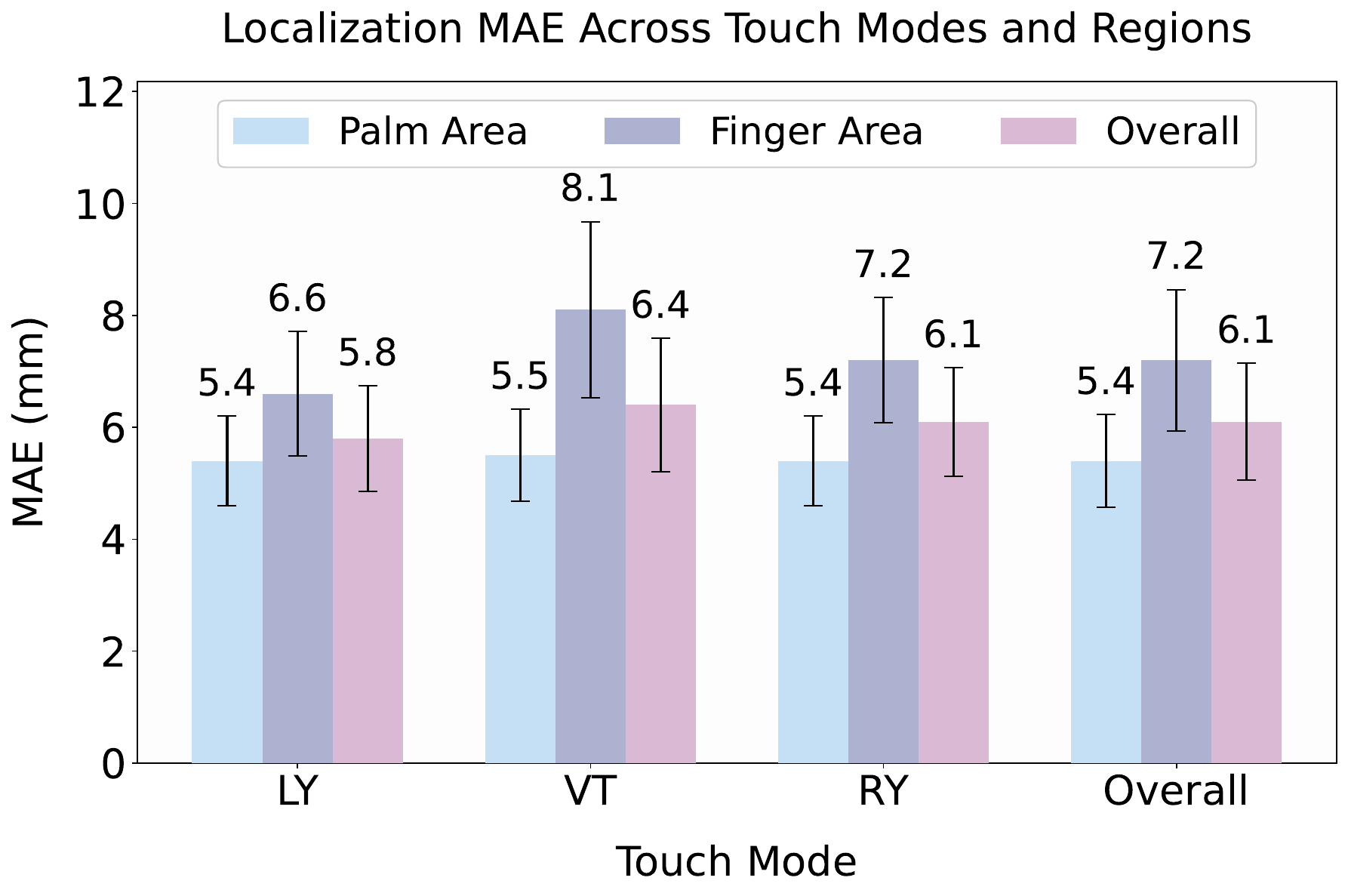}
    \caption{Single-finger MAE by touch mode and region.}
    \label{fig:single_finger_mae}
  \end{subfigure}
  \hfill
  \begin{subfigure}[t]{0.27\textwidth}
    \centering
    \includegraphics[height=2.45cm]{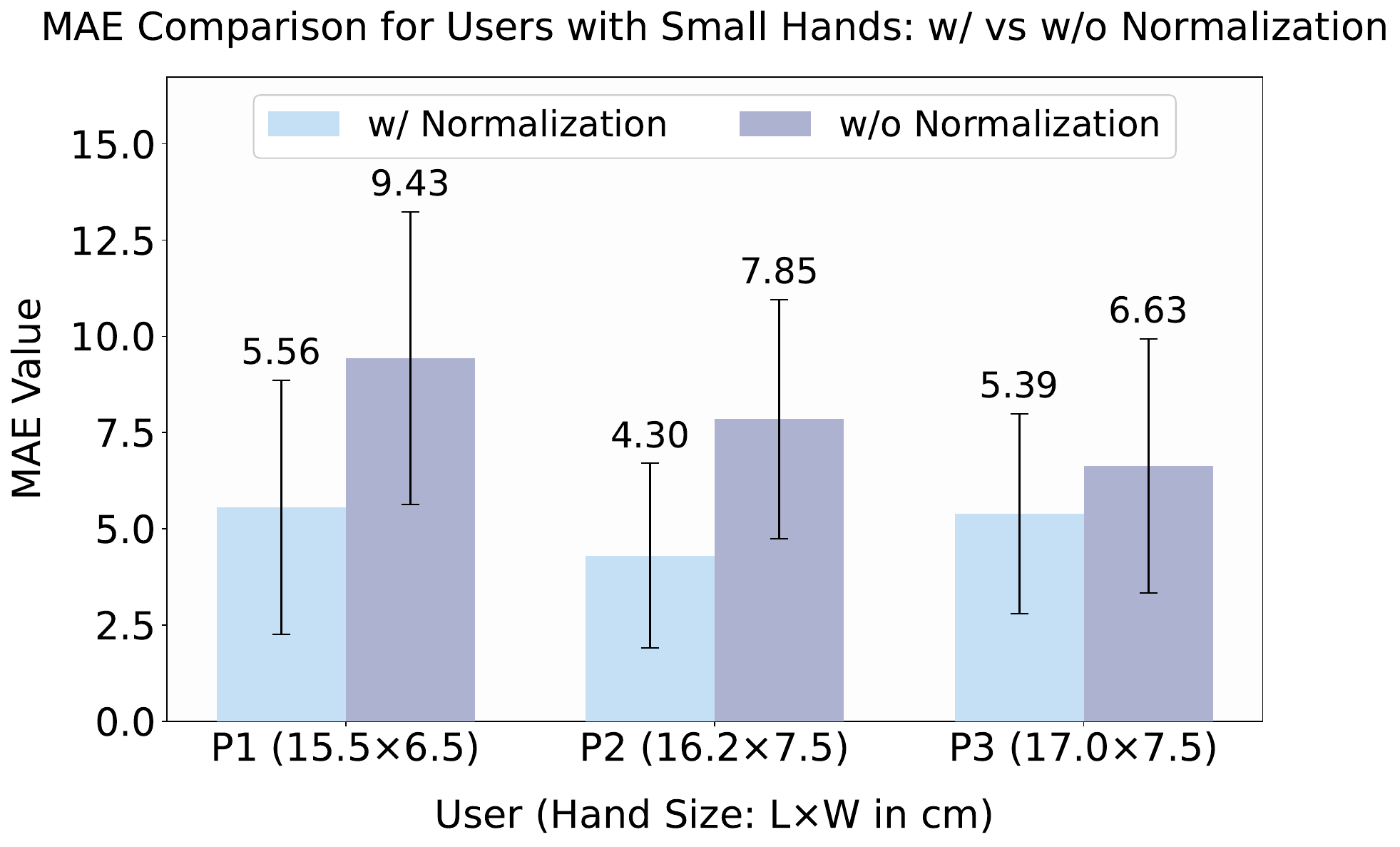}
    \caption{Effect of palm-size normalization.}
    \label{fig:smallhand}
  \end{subfigure}

  \caption{Error analysis and ablation results. Error bars in (c) denote SE; (d) reports the three users with the smallest palms.}
  \label{fig:error_analysis}
  \Description{Four plots summarize localization performance: a two-dimensional error scatter, an error distribution, mean absolute error by touch mode and hand region with standard-error bars, and a comparison showing the effect of palm-size normalization for three small-palm participants.}
\end{figure*}

Under LOPO cross-user evaluation, PalmSpace achieves stable joint performance in touch-state/gesture recognition and absolute localization, reaching 98.9\% binary touch detection accuracy, 96.7\% overall touch gesture F1, and 6.7 mm MAE.

\textbf{Gesture recognition.}
As summarized in Table~\ref{tab:evaluate2}, PalmSpace achieves 98.9\% binary touch detection accuracy. For multi-class gesture recognition, Non-Touch, Single-Touch, Scroll, and Pinch obtain F1 scores of 97.9\%, 97.2\%, 88.1\%, and 97.4\%, respectively, yielding an overall F1 of 96.7\%. 
Among these classes, Scroll is the most challenging due to its similarity to single-finger sliding, yet PalmSpace still achieves an F1 score of 88.1\%.
\textbf{Localization accuracy.}
PalmSpace achieves an overall MAE of 6.7\,mm over all valid touch samples (Table~\ref{tab:evaluate2}). Localization remains accurate for both Single-Touch (6.1\,mm) and Scroll (6.3\,mm), suggesting that the model handles planar finger motion robustly. Even for Pinch, which involves stronger self-occlusion between the thumb and index finger, the error remains within 10.6\,mm under cross-user evaluation.

\textbf{Error patterns.}
For single-finger touch, localization accuracy varies by interaction region. As shown in Fig.~\ref{fig:single_finger_mae}, the Palm Area is easier to localize than the Finger Area (5.4\,mm vs.\ 7.2\,mm MAE). This gap is mainly attributable to larger errors along the $y$-axis in the Finger Area, likely because points farther from the wrist-mounted camera suffer more severe perspective compression. Across touch modes, VT produces the largest error, especially in the Finger Area, whereas LY and RY are more stable.

Fig.~\ref{fig:error_1} and~\ref{fig:error_2} further shows that the prediction errors are tightly concentrated around zero and that most Euclidean errors fall within 1\,cm, indicating good cross-user robustness in practical interaction scenarios. Axis-wise errors are reported in Appendix Table~\ref{tab:evaluate1}.

\begin{table}[ht]
\centering
\caption{Performance overview across different interaction modalities. `Single' refers to single-finger touch.}
\label{tab:evaluate2}

\setlength{\tabcolsep}{4pt}
\resizebox{0.60\columnwidth}{!}{%
\renewcommand{\arraystretch}{0.92}
\begin{tabular}{l c ccc c}
\toprule
\multirow{3}{*}{\textbf{Metric}} & 
\multirow{3}{*}{\textbf{Non-Touch}} & 
\multicolumn{3}{c}{\textbf{Touch Gestures}\textsuperscript{*}} & 
\multirow{3}{*}{\textbf{Overall}} \\ 

\cmidrule(lr){3-5}

 & & \textbf{Single} & \textbf{Scroll} & \textbf{Pinch} & \\ 
\midrule

\textbf{F1 Score (\%)} & 97.9 & 97.2 & 88.1 & 97.4 & 96.7 \\ 
\midrule

\textbf{MAE (mm)} & -- & 6.1 & 6.3 & 10.6 & 6.7 \\ 
\textbf{SD (mm)}  & -- & 4.3 & 3.9 & 11.4 & 5.7 \\ 

\bottomrule
\multicolumn{6}{l}{\scriptsize \textsuperscript{*} The binary Touch Detection Accuracy (Touch vs. Non-Touch) is 98.9\%.} \\
\end{tabular}
}
\end{table}

\definecolor{lightgray}{rgb}{0.92, 0.92, 0.92}

\subsection{Ablation Study}

We assess PalmSpace through indoor LOPO component ablations (Table~\ref{tab:ablation}(a)) and zero-shot evaluation on the unseen daytime data from our Outdoor Robustness Validation (Sec.~\ref{sec:outdoor_study}; Table~\ref{tab:ablation}(b)); its data-collection and training protocol appears in Appendix~\ref{app:outdoor_study}. Unless marked `FT', outdoor models are trained only on indoor data without outdoor fine-tuning.

\textit{In-domain component analysis.}
The full model achieves the best overall indoor performance, with the lowest localization error (6.7 mm MAE) and the highest gesture recognition score (96.7\% F1). 

\textbf{1) Effect of palm-size normalization.} Removing palm-size normalization increases the overall MAE to 6.9 mm, with a drop in multi-finger interactions (8.3 mm to 8.8 mm), indicating improved cross-user generalization, especially for smaller palms. 
This trend is further reflected in the three female participants with the smallest palms in the validation set (Fig.~\ref{fig:smallhand}), whose errors are consistently higher without normalization; for User P2, the MAE nearly doubles.

\textbf{2) Effect of data augmentation.} Removing data augmentation causes the largest drop in gesture recognition, reducing F1 from 96.7\% to 95.3\%, while increasing the overall MAE to 7.0 mm, 
indicating that augmentations simulating camera tilt, camera–palm offset, and illumination variation are important for robustness to pose variation and light changes.

\textit{Out-of-domain architecture and learning analysis.}
Compared with a ResNet-50 encoder, a model without the decoder, and a U-Net baseline, the full PalmSpace architecture achieves the best zero-shot outdoor performance (Table~\ref{tab:ablation}(b)). We further trained two independent copies of the same architecture, one for interaction-state classification and one for touch localization. Compared with unified learning, independent training increases localization MAE from 9.6\,mm to 11.7\,mm and reduces touch accuracy from 93.7\% to 90.5\%, while requiring two deployed model instances. These results suggest that the shared representation improves generalization under distribution shift while reducing runtime model duplication. Outdoor fine-tuning further recovers performance to 6.8\,mm MAE and 98.7\% touch F1.

\begin{table}[ht]
  \centering
  \caption{PalmSpace ablations under (a) in-domain indoor LOPO evaluation and (b) out-of-domain daytime outdoor evaluation. `Independent' uses separately trained interaction-state and localization models; `FT' denotes outdoor fine-tuning.}
  \textbf{(a) In-domain component ablation}\\[2pt]
  \resizebox{0.92\columnwidth}{!}{%
  \renewcommand{\arraystretch}{0.90}
  \begin{tabular}{l|ccc|c}
  \toprule
  \multirow{2}{*}{\textbf{Method}} & \multicolumn{3}{c|}{\textbf{Positioning MAE} ($\downarrow$)} & \multirow{2}{*}{\textbf{Gesture F1} ($\uparrow$)} \\
  \cmidrule(lr){2-4}
   & \textbf{Single-finger} & \textbf{Multi-finger} & \textbf{Overall} & \\
  \midrule
  w/o Normalization      & 6.3 & 8.8 & 6.9 & 96.1 \\
  w/o Data Augmentation  & 6.4 & 8.7 & 7.0 & 95.3 \\
  w/ ResNet50 Encoder    & 6.3 & 8.3 & 6.8 & 96.3 \\
  w/o Decoder            & 6.3 & 8.8 & 6.9 & 96.2 \\
  U-Net                  & 7.0 & 8.5 & 7.4 & 95.8 \\
  \midrule
  \textbf{PalmSpace (Full Model)} & \textbf{6.1} & \textbf{8.3} & \textbf{6.7} & \textbf{96.7} \\
  \bottomrule
  \end{tabular}}\\[5pt]
  \textbf{(b) Out-of-domain architecture and learning ablation}\\[2pt]
  \resizebox{0.92\columnwidth}{!}{%
  \renewcommand{\arraystretch}{0.90}
  \begin{tabular}{l|cc|cc}
  \toprule
  \multirow{2}{*}{\textbf{Method}} & \multicolumn{2}{c|}{\textbf{Positioning}} & \multicolumn{2}{c}{\textbf{Touch}} \\
  \cmidrule(lr){2-3} \cmidrule(lr){4-5}
   & \textbf{MAE} ($\downarrow$) & \textbf{SD} ($\downarrow$) & \textbf{ACC} (\%) ($\uparrow$) & \textbf{F1} (\%) ($\uparrow$) \\
   \midrule
  w/ ResNet50 Encoder & 14.0 & 6.8 & 90.6 & 92.1 \\
  w/o Decoder & 11.9 & 6.1 & 72.5 & 73.1 \\
  U-Net & 17.3 & 8.5 & 89.8 & 91.6 \\
  Independent task-specific models & 11.7 & 7.2 & 90.5 & 93.0 \\
  \midrule
  PalmSpace (Full Model) & \textbf{9.6} & \textbf{5.2} & \textbf{93.7} & \textbf{95.7} \\
  PalmSpace-FT & \textbf{6.8} & \textbf{3.9} & \textbf{98.0} & \textbf{98.7} \\
  \bottomrule
  \end{tabular}
  }
  \label{tab:ablation}
\end{table}

\section{User Study of Single-Finger Interaction}
\label{sec:indooruser}

This section presents an evaluation of PalmSpace for single-finger interaction across both indoor and outdoor settings. 
We first conducted two indoor studies to examine input efficiency under realistic everyday lighting conditions. 
The first focused on pointing and dragging performance using a Fitts' law task, while the second examined eyes-free digit entry performance, with particular attention to the benefits of absolute positioning and proprioception. 
We then assessed the robustness of PalmSpace in representative outdoor conditions, including daytime use under uncontrolled lighting and nighttime use in extremely low-light environments.

\subsection{Input Methods and Participants}

PalmSpace was compared with a smartphone touchscreen baseline, as both support absolute positioning over a similar size of interaction area.
We recruited 12 right-handed participants (3 female; age 19--24, $M=21.2$), all familiar with touchscreen use and none with prior experience using PalmSpace. 
To reflect realistic usage, the indoor study was conducted in 4 indoor settings with different ambient lighting conditions: a morning office, a night meeting room, an evening lab, and an afternoon apartment.
Participants used their dominant index finger for input. For PalmSpace, they used the non-dominant hand without individual calibration; for the touchscreen baseline, they held the smartphone in the non-dominant hand.

\subsection{Study 1: Fitts' Law Study} \label{sec:fittslaw}
\begin{figure*}[tbp]
  \centering
  \begin{subfigure}[t]{0.47\textwidth}
    \centering
    \includegraphics[width=\linewidth]{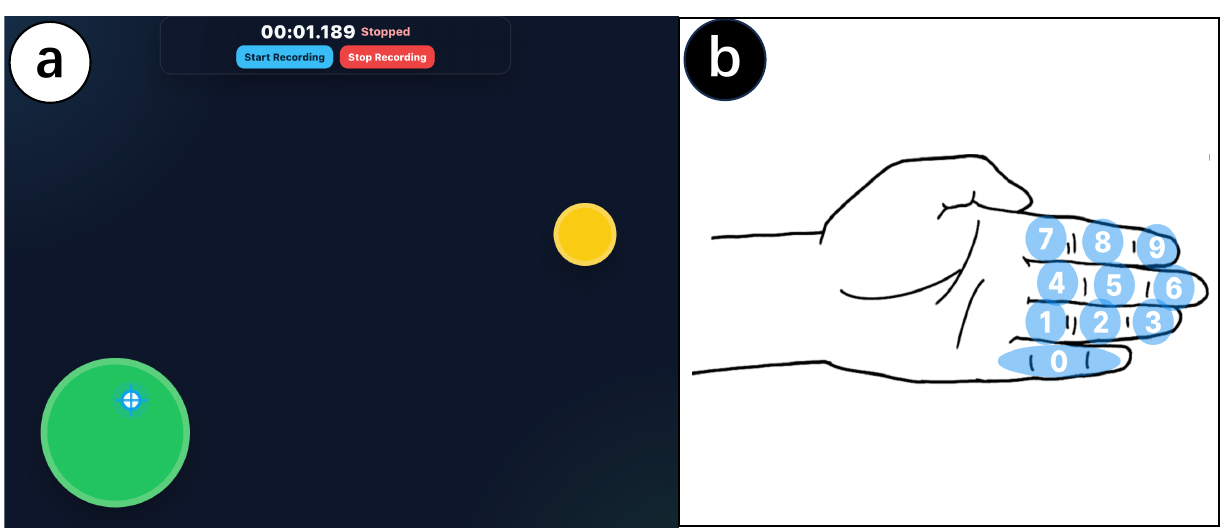}
    \caption{Fitts' law study GUI and PalmSpace numeric keypad layout.}
    \label{fig:user_study_1}
    \label{GUI1:a}
    \label{knuckle_mapping:c}
  \end{subfigure}
  \hfill
  \begin{subfigure}[t]{0.47\textwidth}
    \centering
    \includegraphics[width=\linewidth]{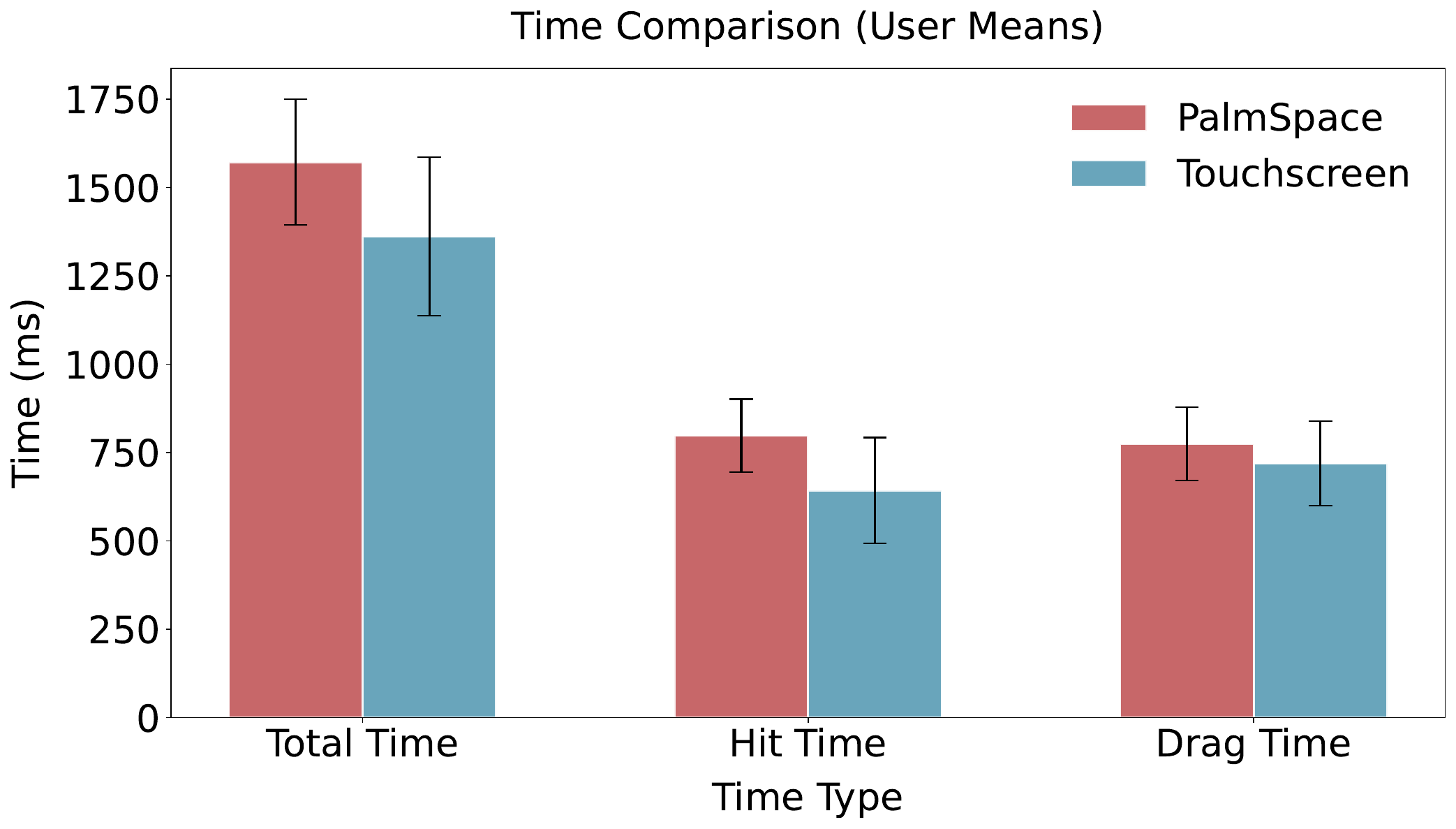}
    \caption{Selection time in the Fitts' law study.}
    \label{fig:fitts_subfig_a}
  \end{subfigure}

  \medskip
  \begin{subfigure}[t]{0.47\textwidth}
    \centering
    \includegraphics[width=\linewidth]{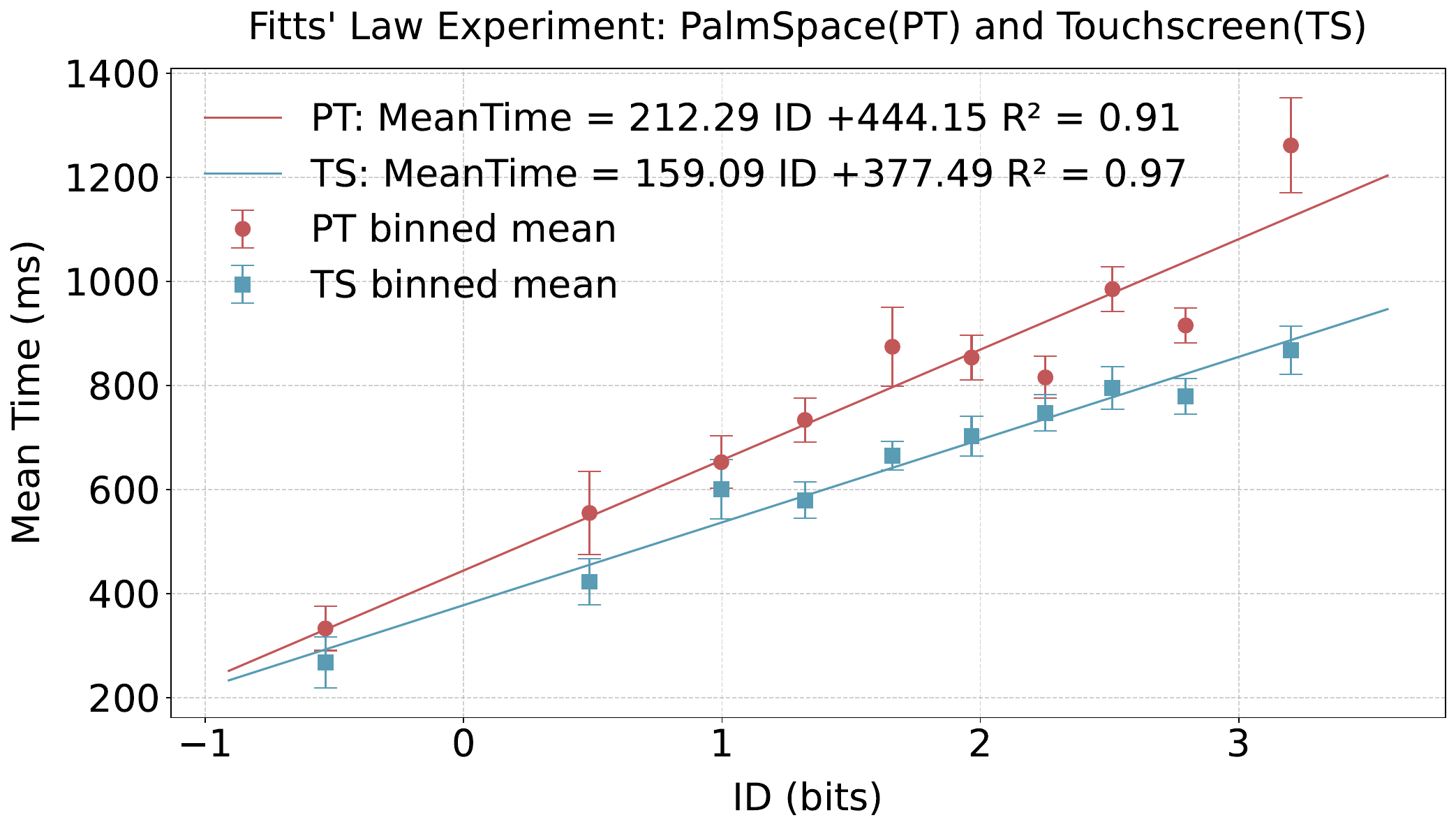}
    \caption{Mean movement time versus index of difficulty.}
    \label{fig:fitts_subfig_b}
  \end{subfigure}
  \hfill
  \begin{subfigure}[t]{0.47\textwidth}
    \centering
    \includegraphics[width=\linewidth]{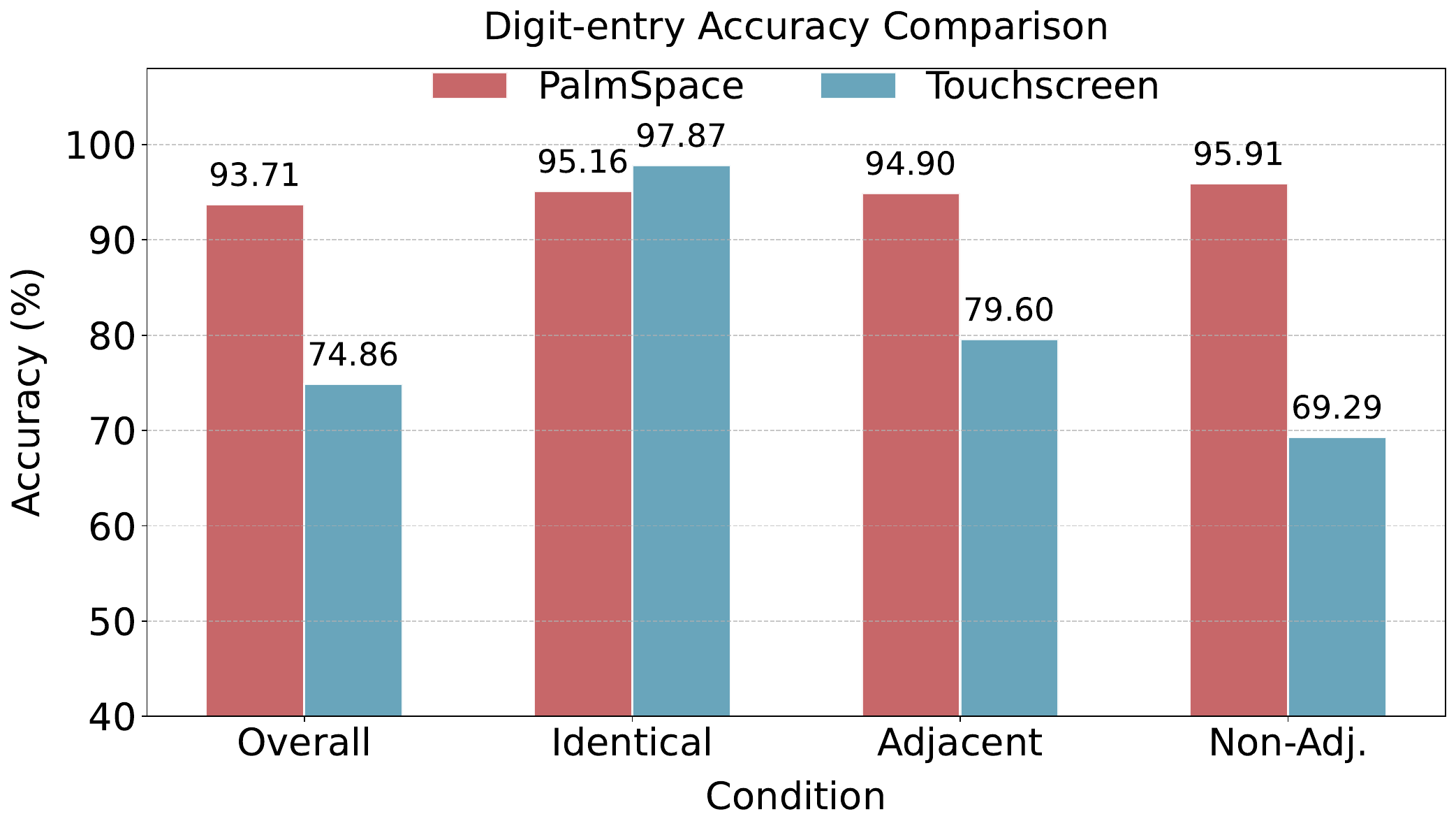}
    \caption{Eyes-free digit-entry accuracy.}
    \label{fig:digit_main}
  \end{subfigure}

  \caption{Setup and results for the indoor single-finger user studies.}
  \label{fig:user_results_combined}
  \Description{A two-by-two layout shows the Fitts' law target-selection interface and palm keypad mapping, selection time by condition, mean movement time by index of difficulty, and eyes-free digit-entry accuracy for PalmSpace and touchscreen input.}
\end{figure*}

\subsubsection{Task and Procedure}
We conducted a Fitts' law study to evaluate pointing and dragging performance. 
In both conditions, interaction was performed with identical visual feedback presented on a separate display (Fig.~\ref{GUI1:a}).
To ensure a fair comparison, the smartphone screen did not display any graphical interface and served only as a coordinate-sensing surface. 
Thus, both conditions used the same interaction mechanism and external visual feedback.
For both methods, normalized input coordinates were mapped to the same GUI, ensuring identical control-display ratios and perceived target sizes across conditions.

In each trial, the GUI displayed a yellow target and a green target area. 
The yellow target had a randomly sampled diameter of 20--40 mm, while the green target area had a fixed diameter of 60 mm. 
The two circles were separated by a random distance of 80--200 mm. 
Participants first moved the cursor to the yellow target. Once selected, the target turned red, indicating the start of the dragging phase. 
Participants then dragged the red circle until it was fully enclosed within the green target area. 
We recorded \textit{hit time}, defined as the time from target onset to target acquisition, and \textit{drag time}, defined as the time from the start of dragging to successful placement.
Each participant completed 5 11-trial blocks per method; excluding the first trial of each block as a warm-up yielded 50 valid trials per method.
The same sequence of target parameters was used across conditions to control task difficulty. 
Participants practiced until performance stabilized, and short breaks were provided between blocks.
\subsubsection{Result} \label{sec:fittsresult}
We analyzed 1,200 valid method-level trials in total (600 per method). 
To verify consistency with Fitts' law, we modeled the relationship between movement time and index of difficulty using $ID=\log_2(\frac{2D}{W})$.
As shown in Fig.~\ref{fig:fitts_subfig_a} and~\ref{fig:fitts_subfig_b}, PalmSpace achieved performance close to the touchscreen baseline. 
For total, hit, and drag times, PalmSpace yielded 1572.2 ms, 797.7 ms, and 774.4 ms, respectively, while the touchscreen yielded 1361.6 ms, 642.5 ms, and 719.2 ms. 
Notably, drag time showed no significant difference between PalmSpace and the touchscreen ($F_{1,11}=1.30, p=0.279$). 
This is particularly surprising because, despite the palm being an uneven and compliant surface, PalmSpace still enabled manipulation performance comparable to a gold-standard touchscreen.

Further analysis of hit time versus ID (Fig.~\ref{fig:fitts_subfig_b}) showed that the main difference arose during target acquisition, likely due to the palm’s less stable surface. 
However, once the target was acquired, dragging performance remained stable and comparable across methods. 
Overall, these results demonstrate that PalmSpace supports efficient pointing and dragging, with performance approaching that of a touchscreen even on the naturally irregular surface of the palm.

\subsection{Study 2: Digit Input Evaluation} \label{sec:digit}

\subsubsection{Task and Procedure}
We evaluated PalmSpace’s support for eyes-free absolute input using a digit entry task. 
Because discrete digit entry requires direct selection of fixed spatial locations, it serves as a natural test of absolute pointing on the palm.

As shown in Fig.~\ref{knuckle_mapping:c}, we mapped twelve finger knuckles (excluding the thumb) to a keypad-like layout, with the three knuckles of the little finger corresponding to digit 0. 
Since the perceived boundaries and absolute positions of knuckle regions can vary slightly across individuals, participants first touched the designated knuckle areas so that their personalized finger layout could be recorded.
This registered the application-specific mapping between anatomical regions and digit labels; it did not retrain or calibrate the PalmSpace sensing model.
For the touchscreen baseline, we used a keypad layout with the same size and spatial arrangement as PalmSpace. 
Participants then completed five rounds of random 10-digit entry under eyes-free conditions. 
The touchscreen baseline followed the same procedure and used the same digit sequences.

\subsubsection{Result}
As shown in Fig.~\ref{fig:digit_main}, PalmSpace significantly outperformed the touchscreen, achieving an overall digit-entry accuracy of 93.71\% versus 74.86\% ($W=2.5, p<.01$). 
We further analyzed transitions between consecutive inputs as \textit{Identical}, \textit{Adjacent}, and \textit{Non-adjacent}, referring to repeated, neighboring, and spatially separated keys, respectively. 
PalmSpace remained stable across all three transition types (all above 94\%), whereas touchscreen performance declined substantially as transition distance increased. 
These results show that PalmSpace effectively supports eyes-free absolute input by leveraging the palm’s intrinsic proprioceptive structure. 
Detailed digit-level results are provided in Appendix~\ref{app:single_finger_details}.

\subsection{Outdoor Robustness Validation}
\label{sec:outdoor_study}

To assess robustness beyond indoor settings, we evaluated PalmSpace in two representative outdoor scenarios: daytime use under uncontrolled lighting and nighttime use in extreme low-light conditions. We recruited 13 new volunteers with no prior experience using the system: 7 performed random palm touches in diverse daytime settings, including parks, streets, and direct sunlight, and 6 performed the same Fitts' law task as in Sec. \ref{sec:fittslaw} at night under ambient illuminance below 5 lux. The daytime model results are reported with the other ablations in Table~\ref{tab:ablation}(b); detailed data-collection and training protocols are provided in Appendix~\ref{app:outdoor_study}.

In daytime testing, the original indoor-trained model achieved a mean absolute error of 9.6 mm; fine-tuning reduced this to 6.8 mm and improved touch detection F1 to 98.7\%, indicating robust accuracy under challenging lighting and background conditions. At night, geometric error could not be measured because the external RGB camera used for ground-truth annotation was ineffective in near-dark environments. We therefore evaluated usability using the same indoor Fitts' law task and task completion time as the primary metric. The average nighttime completion time was 1509.03 ms (SE = 25.49), comparable to the 1572.2 ms observed indoors, suggesting that PalmSpace remains practically usable in near-dark real-world conditions. Overall, these results show that PalmSpace supports efficient single-finger interaction both indoors and across representative outdoor conditions.

\section{Evaluation of Multi-Finger Gestures} \label{sec:multi_eval}

To evaluate PalmSpace in lightweight smart-glasses-oriented scenarios, we conducted a task-based user study on two representative operations in head-worn interfaces: \textit{List Navigation} and \textit{Map Manipulation}. These tasks represent lightweight UI browsing and continuous spatial adjustment, respectively. Accordingly, we evaluated two palm-based multi-finger interactions: \textit{Scroll} for discrete directional navigation and \textit{Pinch} for continuous zooming and rotation.

\begin{figure*}[tbp]
  \centering
  \begin{subfigure}[t]{0.46\textwidth}
    \centering
    \includegraphics[width=0.82\linewidth]{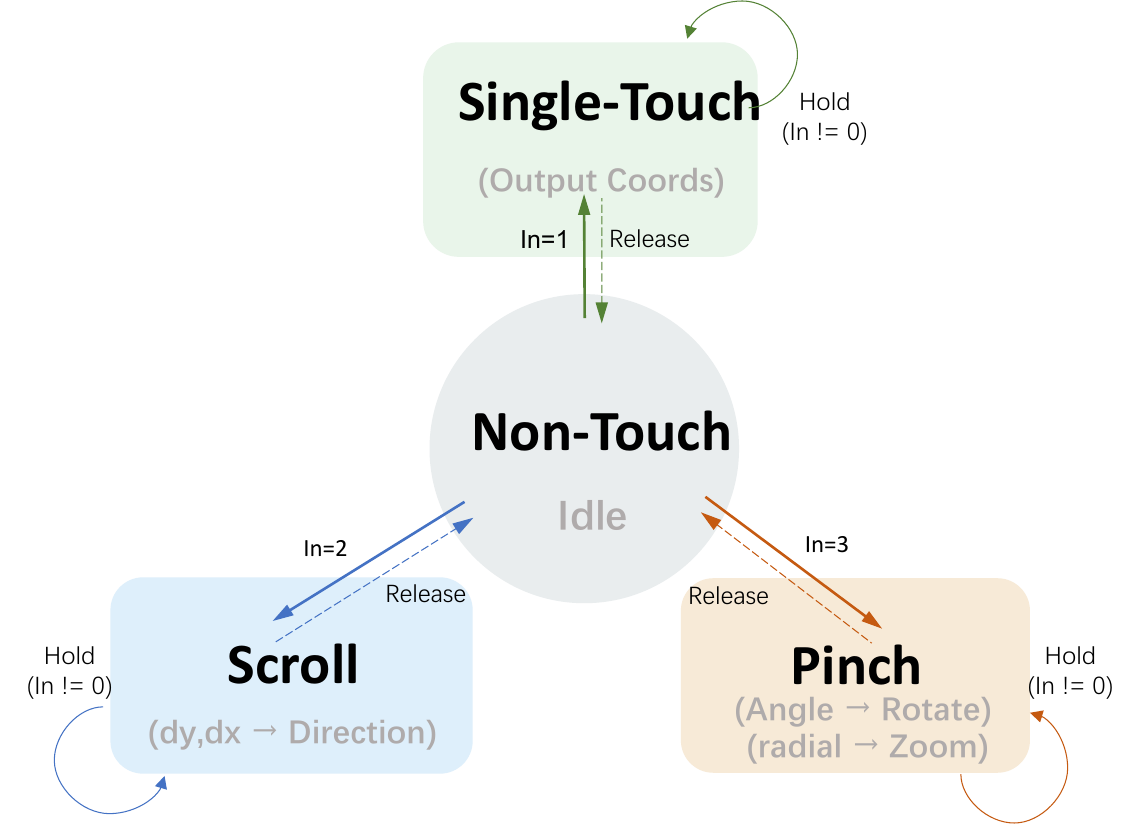}
    \caption{Finite state machine of PalmSpace.}
    \label{fig:state_machine}
  \end{subfigure}
  \hfill
  \begin{subfigure}[t]{0.46\textwidth}
    \centering
    \includegraphics[width=0.78\linewidth]{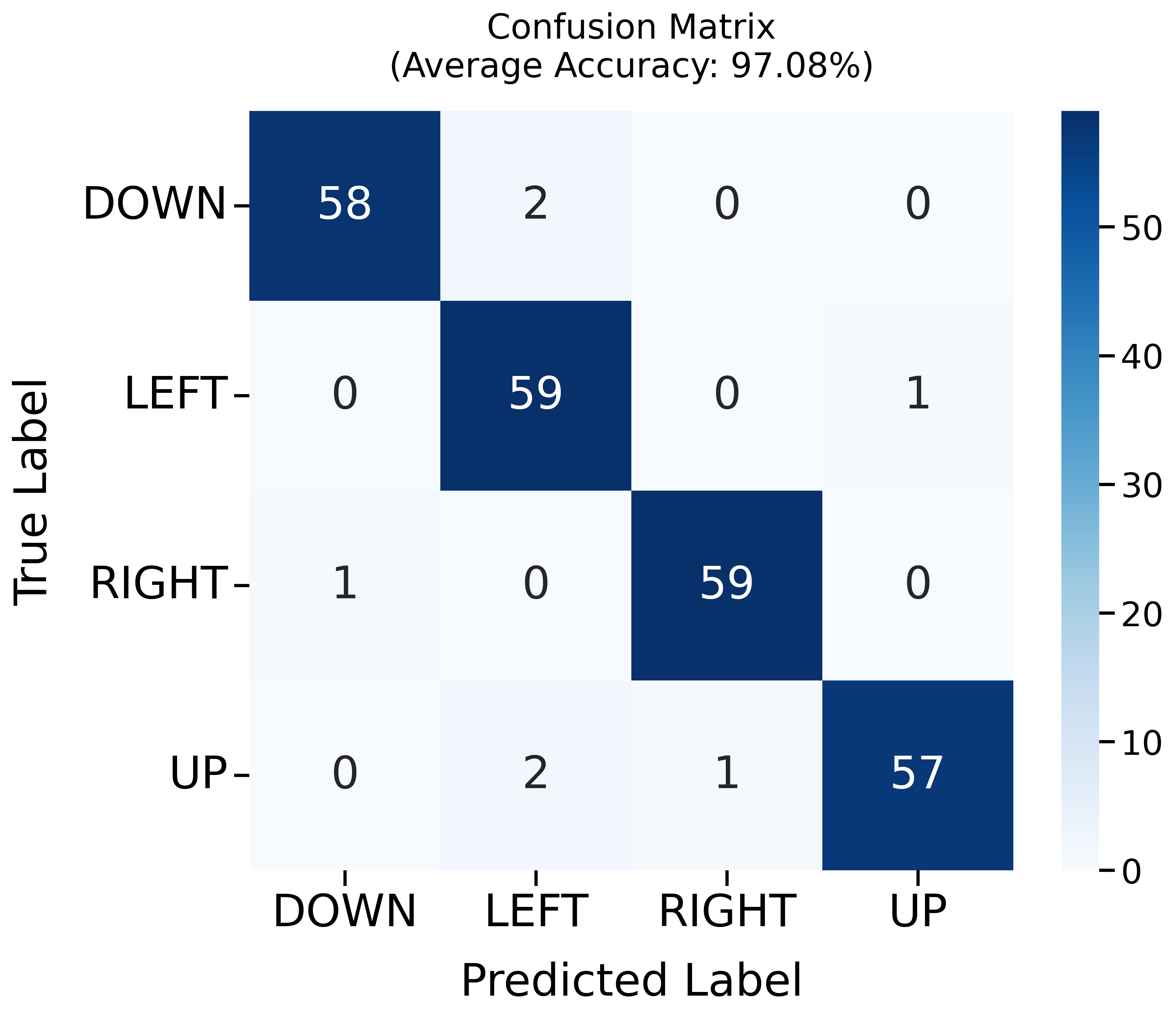}
    \caption{Confusion matrix for the list-navigation task.}
    \label{fig:confusion_matrix}
  \end{subfigure}
  \caption{Interaction logic and recognition results for multi-finger input.}
  \Description{Two panels show PalmSpace's finite state machine and the confusion matrix for four-direction list navigation.}
\end{figure*}

\begin{figure*}[tbp]
  \centering
  \includegraphics[width=0.92\textwidth]{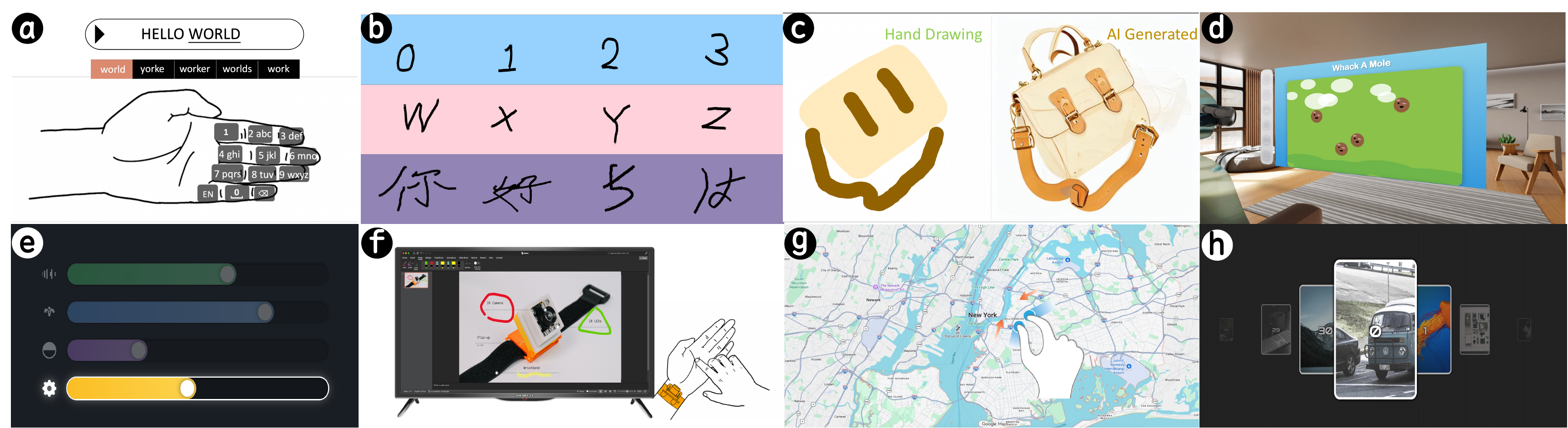}
  \caption{Applications in the PalmSpace design space: (a) T9 keyboard, (b) handwriting, (c) AI-assisted drawing, (d--e) MR controllers, (f) content editing, (g) map manipulation, and (h) page control.}
  \label{fig:apps}
  \setcounter{subfigure}{0}
  \refstepcounter{subfigure}\label{appfig:a}
  \refstepcounter{subfigure}\label{appfig:b}
  \refstepcounter{subfigure}\label{appfig:c}
  \refstepcounter{subfigure}\label{appfig:d}
  \refstepcounter{subfigure}\label{appfig:e}
  \refstepcounter{subfigure}\label{appfig:f}
  \refstepcounter{subfigure}\label{appfig:g}
  \refstepcounter{subfigure}\label{appfig:h}
  \Description{Eight panels show example applications including keypad entry, handwriting, AI-assisted drawing, mixed-reality control, content editing, map manipulation, and page control.}
\end{figure*}

\subsection{Interaction Logic and State Management}
The interaction logic is governed by a finite state machine (FSM), shown in Fig.~\ref{fig:state_machine}. Based on the model output, the system maintains four states: \texttt{Idle}, \texttt{Single-Touch}, \texttt{Scroll Mode}, and \texttt{Pinch Mode}. The system enters the corresponding interaction mode after classification and returns to \texttt{Idle} upon release.
In \textit{Scroll Mode}, directional intent is estimated from the displacement vector between the start and end frames of a gesture sequence. The displacement angle is mapped to one of four directions (\textit{up}, \textit{down}, \textit{left}, \textit{right}), while the displacement magnitude suppresses small unintended motions. This design targets lightweight smart-glasses browsing tasks such as list and menu navigation.
In \textit{Pinch Mode}, the predicted interaction point relative to the palm center is mapped to two continuous control channels: radial displacement for zooming and angular displacement for rotation. This enables compact two-degree-of-freedom control for smart-glasses map manipulation. To keep the main paper concise, the exact transfer functions, gain settings, and angle wrapping compensation are provided in the Appendix. Here, we focus on whether this design provides sufficient controllability, stability, and usability for application-oriented tasks.

\subsection{Experimental Setup}

We recruited 12 participants (age: $M = 25.0$, $SD = 1.87$; balanced gender), all with normal or corrected-to-normal vision. Before the formal study, each participant completed a 5-minute tutorial on gesture triggering and task requirements. Formal trials began only after participants reported being comfortable with the gestures.

\subsubsection{Apparatus and Tasks}
The study comprised two tasks corresponding to the smart-glasses scenarios above. Since current lightweight smart glasses do not support screen projection for customized interfaces, we used an external display and adjusted the viewing distance to approximate smart-glasses display conditions.

\textbf{Task 1: List Navigation.}
This task assessed discrete directional control with \textit{Scroll} gestures. In each trial, participants performed the gesture corresponding to one of four randomized visual prompts (\textit{up}, \textit{down}, \textit{left}, \textit{right}). Each direction was repeated 5 times, yielding 20 trials per user. This task simulates list and menu navigation in smart-glasses interfaces.

\textbf{Task 2: Map Manipulation.}
This task assessed continuous two-degree-of-freedom control with \textit{Pinch} gestures. We used an abstract compound docking task to capture the core demands of smart-glasses map interaction while reducing confounds from rendering latency and scene complexity. The interface showed a semi-transparent \textit{Target State} and a solid \textit{User Cursor}, both rendered as a ``Circle + Arrow'' glyph. Target scale was randomly sampled from $0.5\times$ to $3.0\times$, and target rotation from $0^\circ$ to $360^\circ$. Each participant completed 20 trials, yielding 240 trials in total. Participants aligned the cursor to the target using \textit{Pinch} gestures. A trial was successful only if the cursor stayed within a 10\% scale margin and a $10^\circ$ angular margin for 1.0 s. Scale and rotation were controlled through the same \textit{Pinch} interaction.
We recorded \textit{Completion Time}, \textit{Final Scale Error}, \textit{Final Rotation Error}, and \textit{Overshoots}, where overshoots denote the number of times the controlled value exceeded the target and then reversed for correction during a trial.

\subsection{Results}

For \textit{Task 1} (\textit{List Navigation}), the system achieved an overall recognition accuracy of 97.08\% across 240 trials. As shown in Fig.~\ref{fig:confusion_matrix}, confusion was minimal and evenly distributed, suggesting that \textit{Scroll} provides reliable directional input for list browsing tasks.

For \textit{Task 2} (\textit{Map Manipulation}), participants completed the 240 compound-docking trials in an average of 7.90 s. Rotation stabilized faster than scaling ($T_{rot}=3.59$ s vs. $T_{scale}=4.31$ s), suggesting faster target acquisition in the rotational channel. However, rotation also showed a higher overshoot rate than scaling (0.96 vs. 0.55), indicating more frequent corrective adjustments. In contrast, scaling was slower but more stable. Despite this trade-off, both channels supported precise final control: the final scale error was 4.4\%, and the final rotation error was 6.1$^\circ$, both within the task tolerance. Overall, these results suggest that PalmSpace supports both reliable discrete input and effective continuous zoom-and-rotate control, making it suitable for representative smart-glasses interactions such as list browsing and map manipulation.

\section{Discussion}

\subsection{Usability Feedback}

We assessed PalmSpace’s perceived usability using the System Usability Scale (SUS)~\cite{brooke1996sus}. All participants involved in the user study completed the SUS. Using a 6-point scale, PalmSpace received a mean SUS score of 72.5 $\pm$ 14.0 (median = 70.0, IQR = 16.0), corresponding to a ``Good'' adjective rating on the interpretive scale proposed by Bangor et al.~\cite{bangor2008empirical}. 
Item-level responses suggested strong perceived ease of use, learnability, and confidence, with generally low perceived complexity and awkwardness. 

Interview feedback further contextualized these results. One participant (female, 25) noted, “PalmSpace remained comfortable even during prolonged wear and did not interfere with other ongoing activities. I did not need to pay much attention to my hand, and could stay focused on the screen, which made the experience feel relaxed.” 
Another participant (male, 22) remarked, “The gestures were easy to learn and felt like an almost seamless transition from touchscreen interaction, even though the palm is less flat than touchscreen surface.” Notably, one outdoor-study participant (male, 23) was able to complete the study while mountaineering and camping in snowy conditions, suggesting promising robustness beyond controlled indoor environments.
The results suggest that PalmSpace provides a usable and learnable interaction experience.

\subsection{Design Space}

PalmSpace enables a versatile design space for on-palm interaction, moving beyond isolated touch events toward a more structured interaction paradigm. This design space is defined by the three dimensions of our unified touch modeling framework: contact state, absolute spatial localization, and interaction modality.
Together, these dimensions allow designers to construct a rich vocabulary of eyes-free, calibration-free interactions. 
In doing so, PalmSpace expands the dimensionality and functional scope of prior on-palm input techniques, making the palm an expressive interactive surface.

\textbf{Dimension 1: Contact state.}
It distinguishes spatial exploration from command execution. Non-contact supports hovering and exploration, while contact marks active input. It also supports different temporal forms of interaction, including brief taps and sustained dragging or drawing.

\textbf{Dimension 2: Absolute spatial localization.}
It enables precise selection of coordinates on the palm, supporting both discrete target selection and continuous movement across the palm surface. It can also degrade into relative control, similar to a touchpad.

\textbf{Dimension 3: Interaction modality recognition.}
It distinguishes between single-finger and multi-finger interaction modes, substantially increasing expressive power and enabling more complex compound operations.

We illustrate this design space through three levels of interaction. 
First, combining active contact (Dim. 1), absolute localization (Dim. 2), and single-finger tracking (Dim. 3) supports reliable eyes-free text entry (Sec. \ref{sec:digit}, Fig.~\ref{appfig:a}), handwriting (Appendix \ref{sec:handwriting}, Fig.~\ref{appfig:b}), drawing or editing (Fig.~\ref{appfig:c} and Fig.~\ref{appfig:f}), Whac-A-Mole interaction (Fig.~\ref{appfig:d}), and attribute control such as volume sliders (Fig.~\ref{appfig:e}). 
These examples show how PalmSpace supports symbolic input and precise spatial control through single-finger interaction.
Second, combining active contact (Dim. 1), displacement-based mapping (Dim. 2), and multi-finger tracking (Dim. 3) supports the two representative operations studied in Sec.~\ref{sec:multi_eval}: discrete multi-finger gesture input and continuous compound gesture control, such as zooming, rotation, and sliding. 
These examples demonstrate PalmSpace’s support for manipulation through multi-finger interaction.
Third, PalmSpace also enables \textit{cross-modal composition}, where users combine or transition between single/multi-finger within the same task. 
Such composition supports more complex operations, as illustrated by map interaction (Fig.~\ref{appfig:g}) and page browsing and control (Fig.~\ref{appfig:h}). 
Precise single-finger positioning can be combined with subsequent multi-finger manipulation to accomplish richer workflows. 
Please refer to our supplementary video for all examples.

\subsection{Limitations and Future Work}

Beyond current findings, several natural directions can further extend the system. While the leave-one-participant-out evaluation indicates encouraging robustness across users, the present dataset still captures only part of the variation encountered in real-world hand morphology, appearance, and usage conditions. 
Expanding the participant pool and systematically examining factors such as accessories, fingernail appearance, and a wider range of finger-specific behaviors would help further characterize performance across diverse users and contexts. In addition, the current multi-finger interaction model adopts a robust proxy-based design centered on the primary interaction point. This formulation is well suited to the scrolling and pinch-based tasks studied here, and supports stable, practical control on the compact palm surface. Building on this foundation, future work can explore richer representations of finger articulation and denser hand-surface modeling to support more fine-grained multi-finger interaction in a wider range of tasks.

The current system is implemented as a tethered prototype, a design choice that prioritizes sensing stability and synchronized evaluation in this initial study. This setup enables a controlled validation of the unified touch modeling framework and demonstrates the feasibility of versatile on-palm interaction. Looking ahead, an important next step is to move toward a more compact and self-contained wearable implementation. 
Integrating lightweight cameras into wireless wearables is already a mature direction, as demonstrated by commercial devices such as Pupil Labs Neon \cite{tonsen2020high}, TapXR \cite{tapwithus}, and Wristcam \cite{Wristcam_2020}. 
In a preliminary mobile deployment test, dynamic INT8 quantization and TorchScript reduced model size by 3.7$\times$ and accelerated inference by 2.7$\times$, reaching nearly 30 FPS on a Redmi Turbo 4 smartphone.
Future work will therefore focus on on-device deployment through model compression and system-level optimization, including quantization, pruning, and adaptation to lower-power embedded platforms, while maintaining the interaction quality achieved by the current prototype.
\section{Conclusion}

PalmSpace turns the bare palm into a mode-aware, body-referenced interaction surface by jointly representing touch contact, absolute on-palm location, and interaction mode. Its unified model supports cross-user inference without per-user sensing calibration while avoiding separate state-recognition and localization pipelines. Cross-user evaluation and user studies demonstrate absolute pointing, eyes-free digit input, and representative multi-finger controls such as scrolling and map manipulation, highlighting the potential of unified bare-palm input for future head-worn interaction.
\bibliographystyle{ACM-Reference-Format}
\bibliography{myref}

\appendix

\section{Implementation Details} 
\label{app:system_implementation}

\subsection{Network Configuration}
The visual encoder is initialized from a pre-trained ViT-B/16 model \cite{dosovitskiy2020image}. For single-channel input, the image is replicated to three channels before being fed into the backbone. The backbone output is reshaped into a spatial feature map and projected by a $1\times1$ convolution to a feature dimension of $d_{\text{model}}=256$. Absolute two-dimensional sinusoidal positional embeddings are then added to preserve spatial structure for localization.

The interaction module is a lightweight Transformer composed of three encoder layers. Each layer uses 8 attention heads, a feed-forward expansion ratio of 4, GELU activation, and a dropout rate of 0.1. A learnable class token is prepended to the token sequence, and its output representation is used as the shared feature for downstream prediction. Both the classification and regression heads are implemented as two-layer MLPs with a dropout rate of 0.5. The classification branch outputs four logits corresponding to \textit{Non-contact}, \textit{Single-finger}, \textit{Scroll}, and \textit{Pinch}, while the regression branch predicts the normalized 2D interaction coordinates $(x,y)$.

The deployed model contains approximately 89M parameters, striking a practical balance between representational capacity and inference efficiency. Despite jointly solving three tasks---touch contact detection, gesture classification, and absolute position regression---the model runs efficiently on a Mac mini with an M4 Pro chip used in our user study, achieving an average inference latency of 7.2\,ms per frame. This corresponds to a theoretical throughput of over 100\,Hz, which is sufficient to fully match the refresh rate of the current camera pipeline. Therefore, the model inference stage does not constitute the primary bottleneck of the system. Instead, it provides enough computational headroom for stable real-time interaction, ensuring that continuous palm-writing and multi-finger manipulations can be processed with low perceived delay. These findings support the practicality of our model design for deployment in interactive, low-latency scenarios.
\subsection{Data Augmentation}
To improve robustness under practical use conditions, we apply data augmentation to simulate three common sources of variation in wrist-worn deployment. First, because the wrist-mounted camera may be slightly tilted during wearing, input images are randomly rotated within $[-10^\circ, 10^\circ]$. Second, because the relative placement between the camera and the bottom of the palm may vary, we apply random translation with a maximum relative offset of 0.1 in both horizontal and vertical directions. Third, because ambient illumination may change across environments, we randomly adjust image brightness and contrast with an offset factor of 0.2. These augmentations improve robustness to small pose perturbations and lighting variation in real-world use.

\subsection{Training Details} 
All experiments are conducted on an NVIDIA GeForce RTX 4090 GPU with 24GB VRAM. The original image resolution is $1920\times1080$, and all images are resized to $224\times224$ for training and inference. The model is trained using the Adam optimizer with an initial learning rate of $1\times10^{-4}$, weight decay of $1\times10^{-4}$, and batch size of 160. Training runs for up to 100 epochs with ReduceLROnPlateau for adaptive learning-rate scheduling and early stopping to mitigate overfitting.

\subsection{System Workflow}
\begin{figure}[ht]
  \centering
  \includegraphics[width=0.74\linewidth]{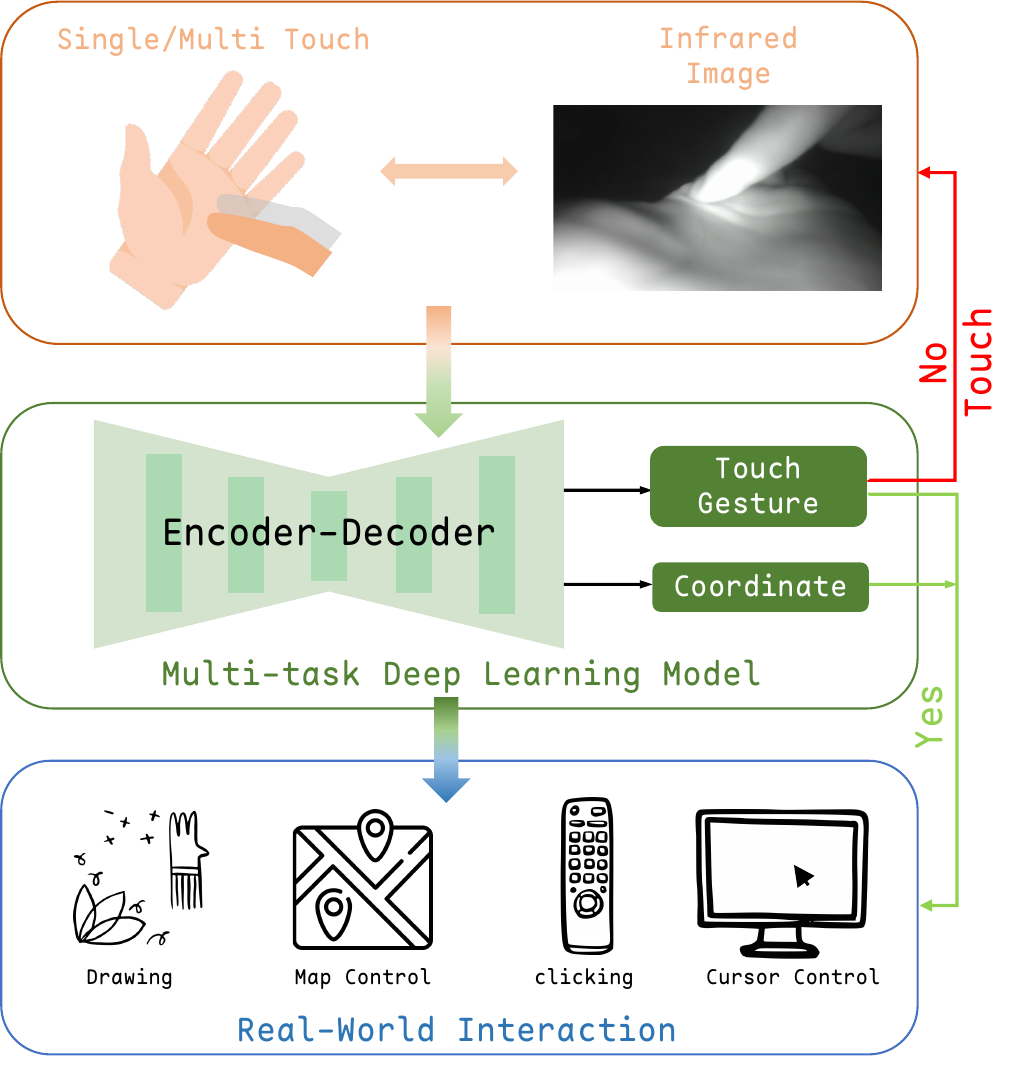}
  \caption{System workflow: The user interacts by touching their non-dominant palm with their dominant hand's finger. A deep learning model processes this input in real-time, predicting both touch events and precise touch positions. Upon detecting valid touch interactions, the system maps these absolute coordinates to specific applications in the real world.}
  \label{fig:workflow}
  \Description{A left-to-right workflow shows infrared palm capture, neural-network inference of touch state, gesture class and location, and mapping of those predictions to single-finger drawing or multi-finger application commands.}
\end{figure}

Fig. \ref{fig:workflow} illustrates the workflow: the user wears a wristband that illuminates the palm with infrared light and performs interactions using either a single finger or multiple fingers. Captured images are processed by a multi-task Transformer network to simultaneously predict the touch contact state, gesture class (i.e., single-finger, scroll, or pinch), and the absolute position. When contact is detected, the system applies different interaction logic based on the gesture class: for single-finger mode, the system renders points on the canvas to simulate a realistic writing experience; for multi-finger gestures, the system maps the predicted coordinates to scrolling or zooming commands, enabling continuous spatial manipulation.

\begin{figure}[ht]
  \centering
  \includegraphics[width=0.4\linewidth]{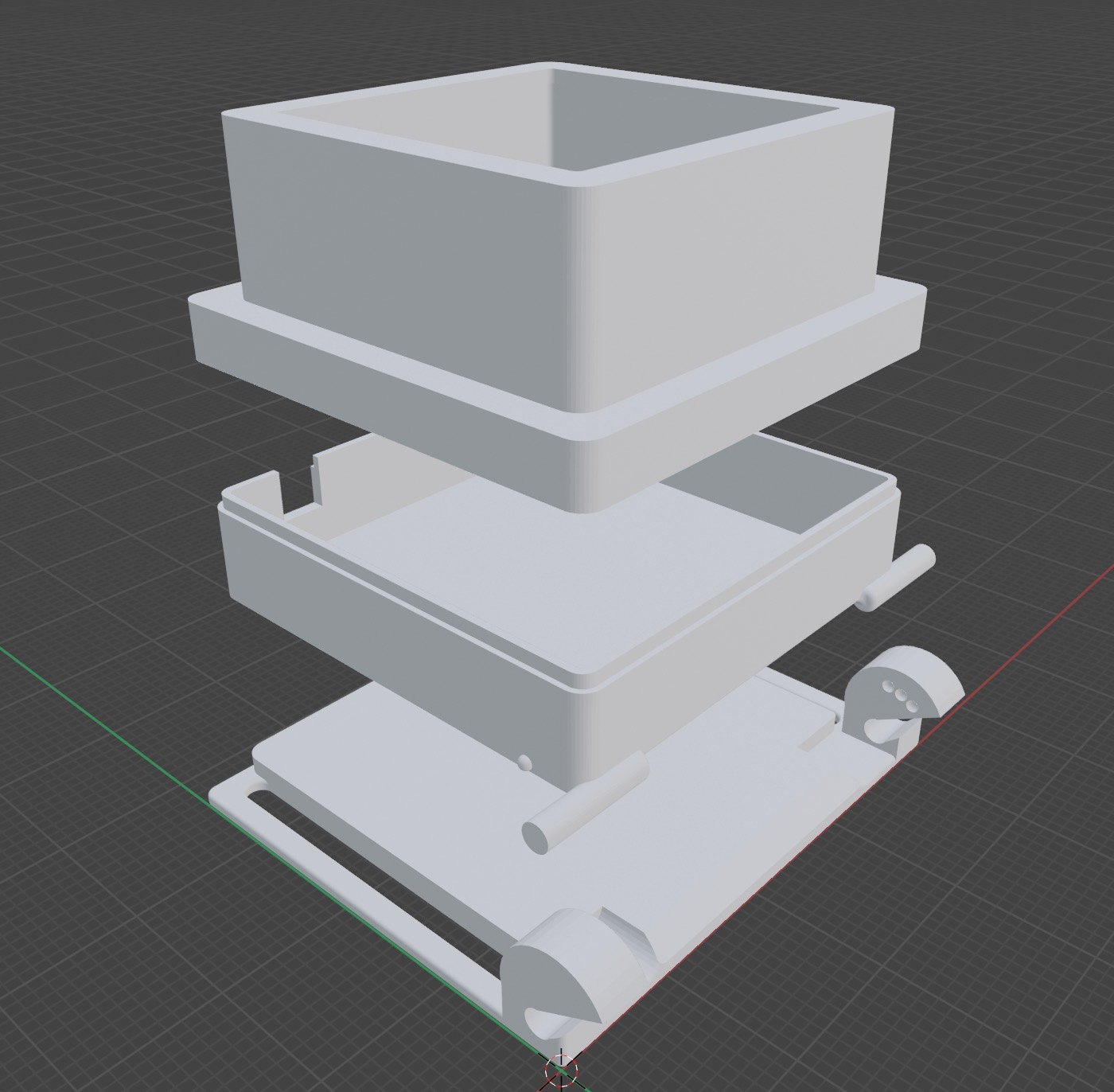}
  \caption{Blender render of the wristband casing.}
  \label{fig:3dmodel}
  \Description{A three-dimensional rendering shows the compact rectangular wristband casing and its hinged camera mount.}
\end{figure}

\section{Data Collection Details}
\label{app:datacollect}

\begin{figure}[ht]
  \centering
  \includegraphics[width=0.8\linewidth]{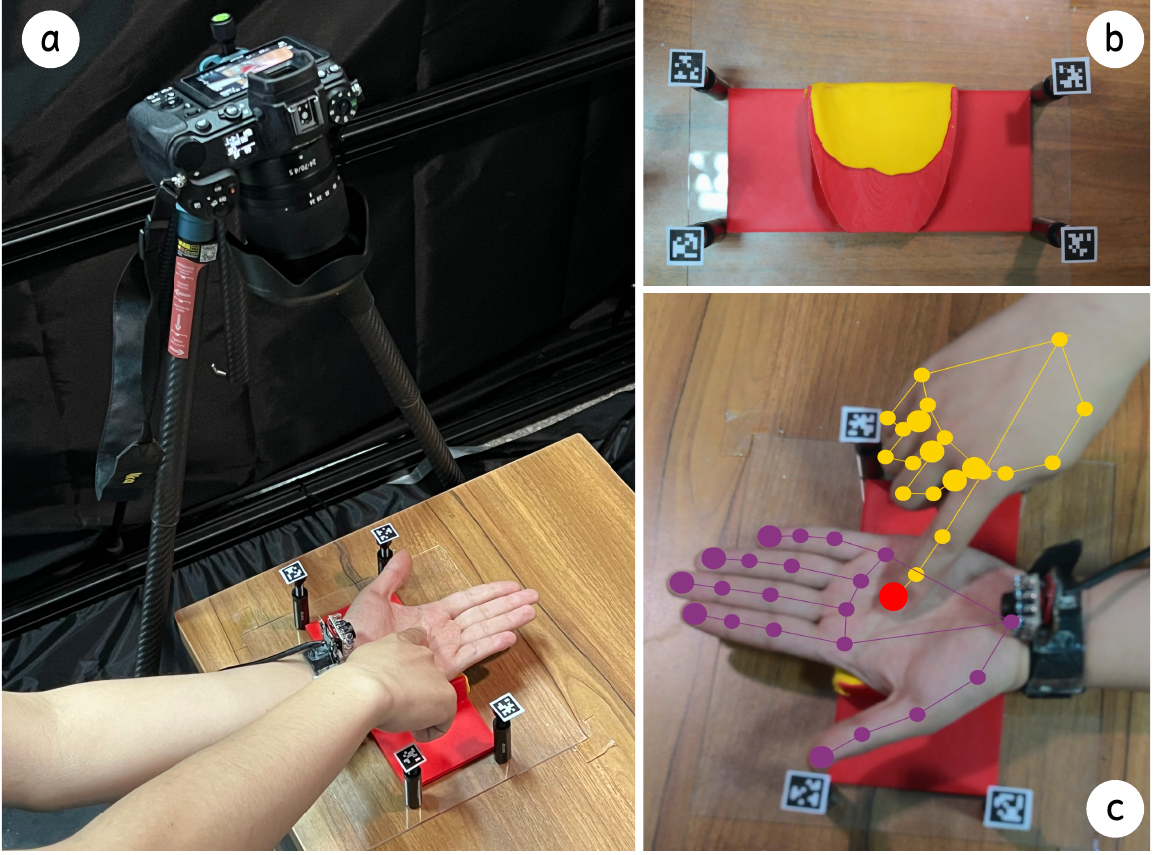}
  \caption{Data collection setup: (a) Overall architecture of the capture system; (b) 3D-printed calibration device and corresponding calibration system; (c) Tracking of the dominant hand's index fingertip pixel coordinates via MediaPipe \cite{zhang2020mediapipe} and their transformation into the calibrated palm plane coordinate system.}
  \label{fig:apparatus}
  \setcounter{subfigure}{0}
  \refstepcounter{subfigure}\label{subfig:a}
  \refstepcounter{subfigure}\label{subfig:b}
  \refstepcounter{subfigure}\label{subfig:c}
  \Description{Three panels show the complete data-capture setup, the three-dimensional printed calibration fixture and calibration interface, and fingertip tracking transformed from camera pixels into coordinates on the palm plane.}
\end{figure}

\subsection{Apparatus and Calibration}
Fig. \ref{subfig:a} illustrates the data collection setup, consisting of a wrist-mounted IR camera (input) and an overhead Nikon Z6 II camera (ground truth) equipped with a Nikkor 24--70 mm f/4S lens. The overhead camera streams 4K video at 30 FPS to a Mac Mini M4 Pro and is used only for calibration and ground-truth generation during data collection, rather than in deployment.

To align the palm surface with the calibration plane, we used a 3D-printed groove filled with molding clay (Fig. \ref{subfig:b}) to accommodate different hand shapes. Four height-adjustable posts marked with AprilTags \cite{apriltag} define the physical coordinate system within a $19.8 \times 8.5$ cm rectangular area while avoiding marker occlusion in the wrist camera's field of view.

We compute a homography matrix $H$ to map the overhead camera's pixel coordinates to the physical calibration plane. Using the four AprilTags, we estimate $H$ via least squares. Given a pixel coordinate $(u, v)$, the corresponding physical position $(x, y)$ is obtained as:
\begin{equation}
\begin{bmatrix} x' \\ y' \\ z' \end{bmatrix} = H \begin{bmatrix} u \\ v \\ 1 \end{bmatrix}, \quad
x = \frac{x'}{z'}, \quad
y = \frac{y'}{z'}.
\end{equation}
We approximate the palm as a planar surface; due to the large distance between the overhead camera and the hand, the effect of minor palm undulations is negligible.

\subsection{Ground Truth Generation}
Fingertip positions are tracked using Google's MediaPipe \cite{zhang2020mediapipe}, a robust keypoint detection method widely used in hand pose research \cite{kyu2024eitpose,lee2024echowrist,devrio2022discoband}. The detected pixel coordinates (red marker in Fig. \ref{subfig:c}) are transformed into physical coordinates through $H$. Compared with reflective-marker-based optical tracking systems \cite{dezfuli2012palmrc,wang2015palmgesture}, this markerless approach avoids visual artifacts in the wrist-mounted IR images and better preserves realistic bare-hand interaction conditions.

A Python script running on the Mac Mini M4 Pro synchronizes the two cameras at 15 FPS and computes ground-truth positions in real time. Before each experiment, we calibrated the system and ensured a reprojection error below $10^{-5}$.

\subsection{Collection Procedure}
Data collection was conducted in quiet indoor settings at different times of day (morning, afternoon, and evening), with the devices placed in different locations within the room. The study included 9 sessions: 6 single-finger sessions (2 areas $\times$ 3 modes), 2 multi-finger sessions (Scroll and Pinch), and 1 non-contact session.

For single-finger sessions, participants performed serpentine sliding patterns. Interactions in the palm area included both longitudinal and transverse motions, whereas the finger area was limited to longitudinal sliding because of finger gaps. Unlike PalmGesture \cite{wang2015palmgesture}, we allowed participants to adopt comfortable finger pitch angles. For the two multi-finger sessions, interactions were restricted to the palm surface. To preserve ecological validity, we did not impose strict pose constraints; instead, participants performed gestures according to their natural habits, mimicking common touchpad interactions. The non-contact session collected negative samples such as hovering and random gestures to help distinguish valid touches from false positives.

We explicitly excluded force-sensing resistors (FSRs) \cite{mollyn2024egotouch} for ground-truth collection. In pilot tests, FSRs introduced visual occlusion and were insufficiently sensitive to the light-touch interactions typical in our scenarios. To ensure coordinate consistency, participants aligned their palms within the 3D-printed groove relative to the AprilTag markers, with the palm center defined as the origin. During capture, users wore the device without an outer case to avoid occlusion and were encouraged to flex their wrists within an approximate range of $\pm 30^\circ$ to increase pose diversity while maintaining palm-plane visibility.

\section{Supplementary Details for Single-Finger Study}
\label{app:single_finger_details}

\begin{table*}[htbp]
  \centering
  \caption{The average MAE and SD in different interaction modes and touch regions. Errors are reported in mm.}
  \renewcommand{\arraystretch}{0.9}
  \setlength{\tabcolsep}{9pt}
  \resizebox{0.99\textwidth}{!}{%
  \begin{tabular}{llccc ccc ccc}
    \toprule
    & & \multicolumn{3}{c}{\textbf{Palm-Area}} & \multicolumn{3}{c}{\textbf{Finger-Area}} & \multicolumn{3}{c}{\textbf{Overall}} \\
    \cmidrule(lr){3-5} \cmidrule(lr){6-8} \cmidrule(lr){9-11}
    \textbf{Mode} & \textbf{Metric} & 
    \textbf{$x_{\text{error}}$} & \textbf{$y_{\text{error}}$} & \textbf{$l_{\text{error}}$} & 
    \textbf{$x_{\text{error}}$} & \textbf{$y_{\text{error}}$} & \textbf{$l_{\text{error}}$} & 
    \textbf{$x_{\text{error}}$} & \textbf{$y_{\text{error}}$} & \textbf{$l_{\text{error}}$} \\
    \midrule
    \rowcolor{lightgray}
    \textbf{LY} & MAE & 2.8 & 3.9 & 5.4 & 3.5 & 5.0 & 6.6 & 3.1 & 4.4 & 5.8 \\
    \rowcolor{lightgray}
    & SD  & 2.6 & 3.1 & 3.3 & 3.2 & 4.3 & 4.6 & 2.9 & 3.6 & 3.9 \\
    \midrule
    \textbf{VT} & MAE & 3.1 & 4.1 & 5.5 & 4.5 & 5.8 & 8.1 & 3.6 & 4.7 & 6.4 \\
    & SD  & 2.9 & 3.0 & 3.4 & 4.6 & 5.6 & 6.5 & 3.6 & 4.2 & 4.9 \\
    \midrule
    \rowcolor{lightgray}
    \textbf{RY} & MAE & 2.8 & 3.9 & 5.4 & 3.8 & 5.4 & 7.2 & 3.2 & 4.5 & 6.1 \\
    \rowcolor{lightgray}
    & SD  & 2.5 & 3.2 & 3.3 & 3.2 & 4.3 & 4.6 & 2.9 & 3.8 & 4.0 \\
    \midrule
    \textbf{Overall} & MAE & 2.9 & 3.9 & 5.4 & 3.9 & 5.4 & 7.2 & 3.3 & 4.5 & 6.1 \\
    & SD  & 2.7 & 3.1 & 3.4 & 3.7 & 4.7 & 5.2 & 3.1 & 3.8 & 4.3 \\
    \bottomrule
  \end{tabular}
  }
  \label{tab:evaluate1}
\end{table*}

\subsection{Indoor Settings}
To avoid restricting the evaluation to a single controlled environment, the indoor study was distributed across four everyday real-world scenarios: a morning office (4 users), a night meeting room (2), an evening lab (2), and an afternoon apartment (4).

\subsection{Additional Details of the Digit-Entry Study}
\label{app:digit}
For digit entry, twelve finger knuckles (excluding the thumb) were mapped to a keypad-like layout, with the three knuckles of the little finger representing digit 0. 
Participants first sequentially touched the designated knuckle regions. We recorded the corresponding coordinates, discarded the first and last 10\% of samples to reduce noise, and used the remaining samples to construct a 3-nearest-neighbor classifier for knuckle identification. This stage established only the mapping between anatomical locations and digit labels and did not calibrate the PalmSpace sensing model itself.

In the testing phase, participants completed five rounds of random 10-digit sequences under eyes-free conditions. The touchscreen baseline followed the same protocol, with the right half of the screen partitioned into ten digit-mapped zones and visual feedback provided during registration to facilitate layout learning.

\begin{figure}[ht]
  \centering
  \begin{subfigure}[b]{0.48\textwidth}
    \centering
    \includegraphics[width=\textwidth]{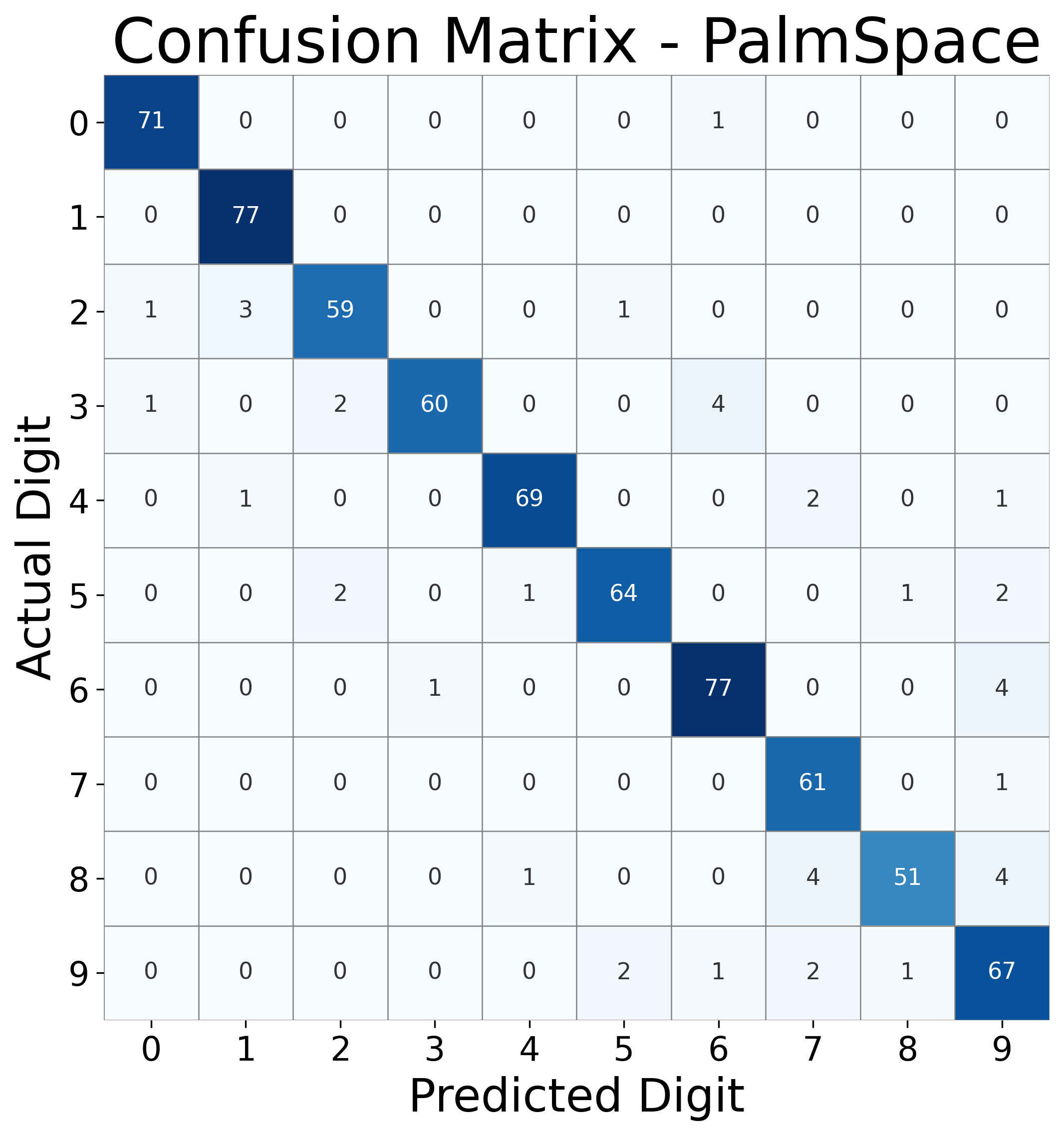}
    \caption{PalmSpace}
    \label{fig:cm_PalmSpace}
  \end{subfigure}
  \hfill
  \begin{subfigure}[b]{0.48\textwidth}
    \centering
    \includegraphics[width=\textwidth]{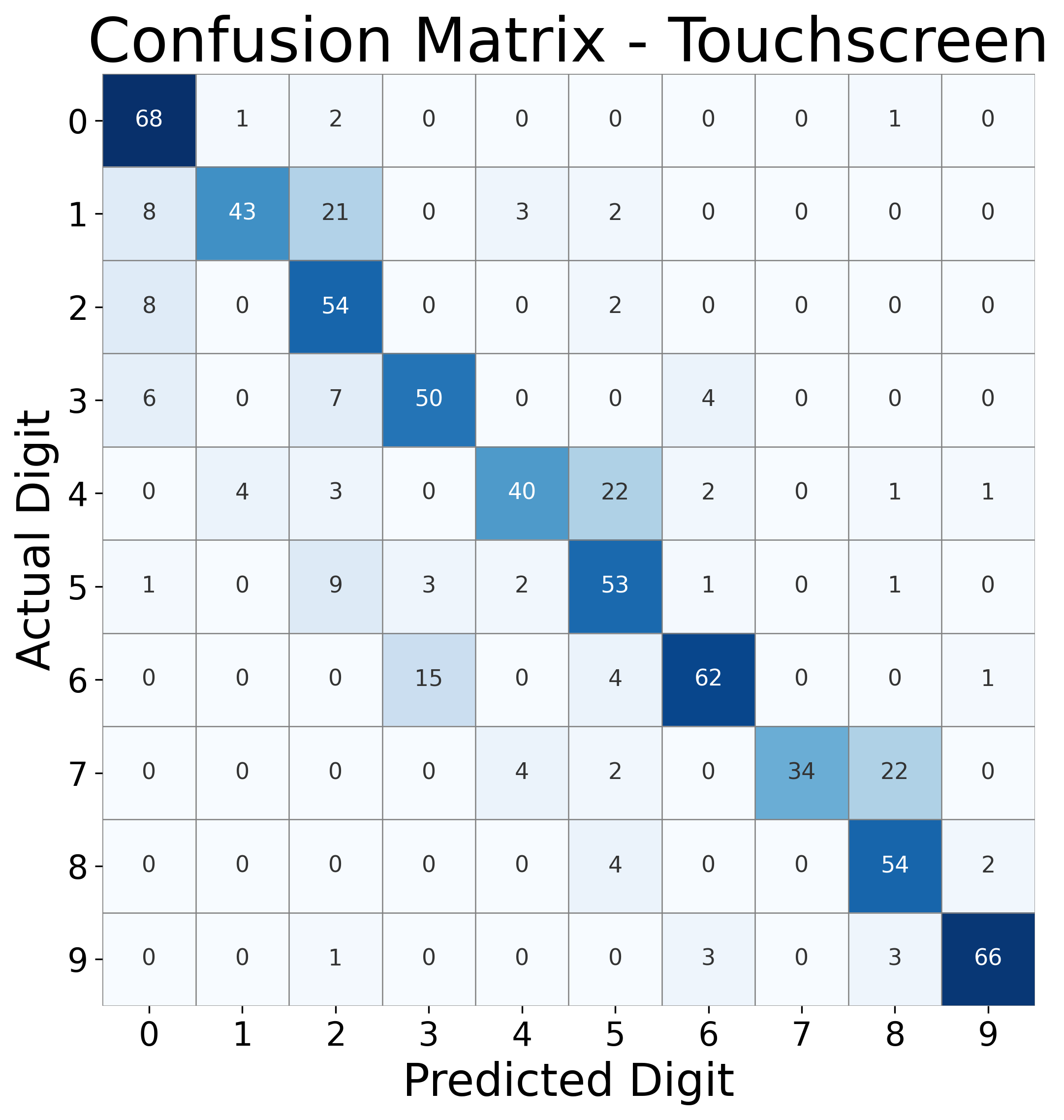}
    \caption{Touchscreen}
    \label{fig:cm_touchscreen}
  \end{subfigure}
  \caption{Confusion matrices for digit input: (a) PalmSpace, (b) Touchscreen.}
  \label{fig:cm_comparison}
  \Description{Side-by-side ten-class confusion matrices compare intended and recognized digits for PalmSpace and the touchscreen baseline. PalmSpace errors are more evenly distributed, whereas touchscreen errors are concentrated among several neighboring digits.}
\end{figure}

The detailed digit-level and transition-level results are shown in Table \ref{tab:accuracy_breakdown}, and the confusion matrices are shown in Fig. \ref{fig:cm_comparison}. The largest touchscreen errors occurred for digits such as 1, 4, and 7, whereas PalmSpace maintained more uniform performance across digits. Transition analysis further showed that touchscreen accuracy dropped substantially from identical to adjacent and non-adjacent transitions, while PalmSpace remained stable across all three categories.

\begin{table*}[t]
  \centering
  \caption{Comparison of Digit Recognition and Transition Accuracy (\%) between PalmSpace and Touchscreen.}
  \label{tab:accuracy_breakdown}
  \resizebox{\textwidth}{!}{%
  \begin{tabular}{lccccccccccccc}
    \toprule
    & \multicolumn{10}{c}{\textbf{Digit Recognition Accuracy (\%)}} & \multicolumn{3}{c}{\textbf{Transition Accuracy (\%)}} \\
    \cmidrule(lr){2-11} \cmidrule(lr){12-14}
    \textbf{Method} & \textbf{0} & \textbf{1} & \textbf{2} & \textbf{3} & \textbf{4} & \textbf{5} & \textbf{6} & \textbf{7} & \textbf{8} & \textbf{9} & \textbf{Identical} & \textbf{Adjacent} & \textbf{Non-Adj.} \\
    \midrule
    PalmSpace   & 98.55 & 100.00 & 90.74 & 89.47 & 95.16 & 91.94 & 97.06 & 100.00 & 92.31 & 95.00 & 95.16 & 94.90 & 95.91 \\
    Touchscreen & 95.65 & 46.77  & 85.19 & 70.18 & 54.84 & 74.19 & 75.00 & 48.15  & 90.38 & 88.33 & 97.87 & 79.60 & 69.29 \\
    \bottomrule
  \end{tabular}%
  }
\end{table*}

\subsection{Handwriting Task User Study} 
\label{sec:handwriting}
Handwritten character input requires extremely high accuracy in finger touch detection and absolute finger positioning, as even small deviations can render text visually unrecognizable. Therefore, we conducted a handwriting input experiment to assess continuous input efficiency.

\subsubsection{Task and Procedure}
We selected the 10 digits (0-9) and 26 uppercase letters (A-Z) as input tasks. Each participant was randomly assigned 20 characters to write sequentially following GUI prompts, using their preferred palm posture. Thanks to advancements in VLM (Vision Language Model) technology, instead of training a dedicated model, we employed GPT-4o \cite{openai2024gpt4o} for recognition. We used GPT-4o for its robustness to scale/rotation and direct structured output, unlike OCR models (e.g., Tesseract \cite{TesseractOCR} and Paddle \cite{PaddleOCR}), which are sensitive to scale/rotation and need complex tuning. For the touchscreen, participants rewrote the same 20 characters. In total, 12 $\times$ 2 $\times$ 20 = 480 characters were collected.

\subsubsection{Results}
\begin{figure}[htbp]
  \centering
  \begin{subfigure}[t]{0.62\textwidth}
    \centering
    \includegraphics[width=\linewidth]{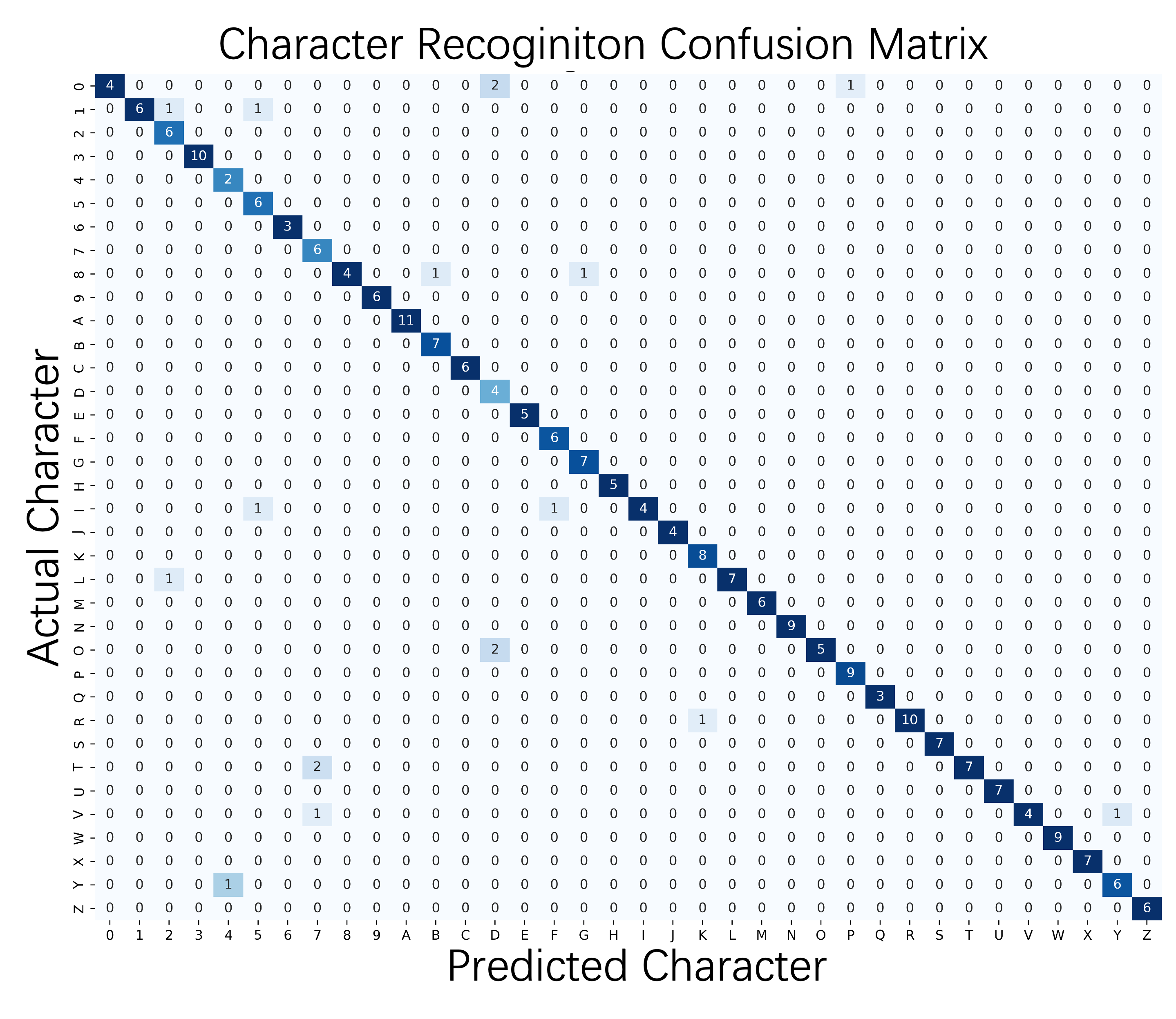}
    \caption{Confusion matrix for 36 handwritten characters.}
    \label{fig:hand_subfig_a}
  \end{subfigure}
  \hfill
  \begin{subfigure}[t]{0.32\textwidth}
    \centering
    \includegraphics[width=\linewidth]{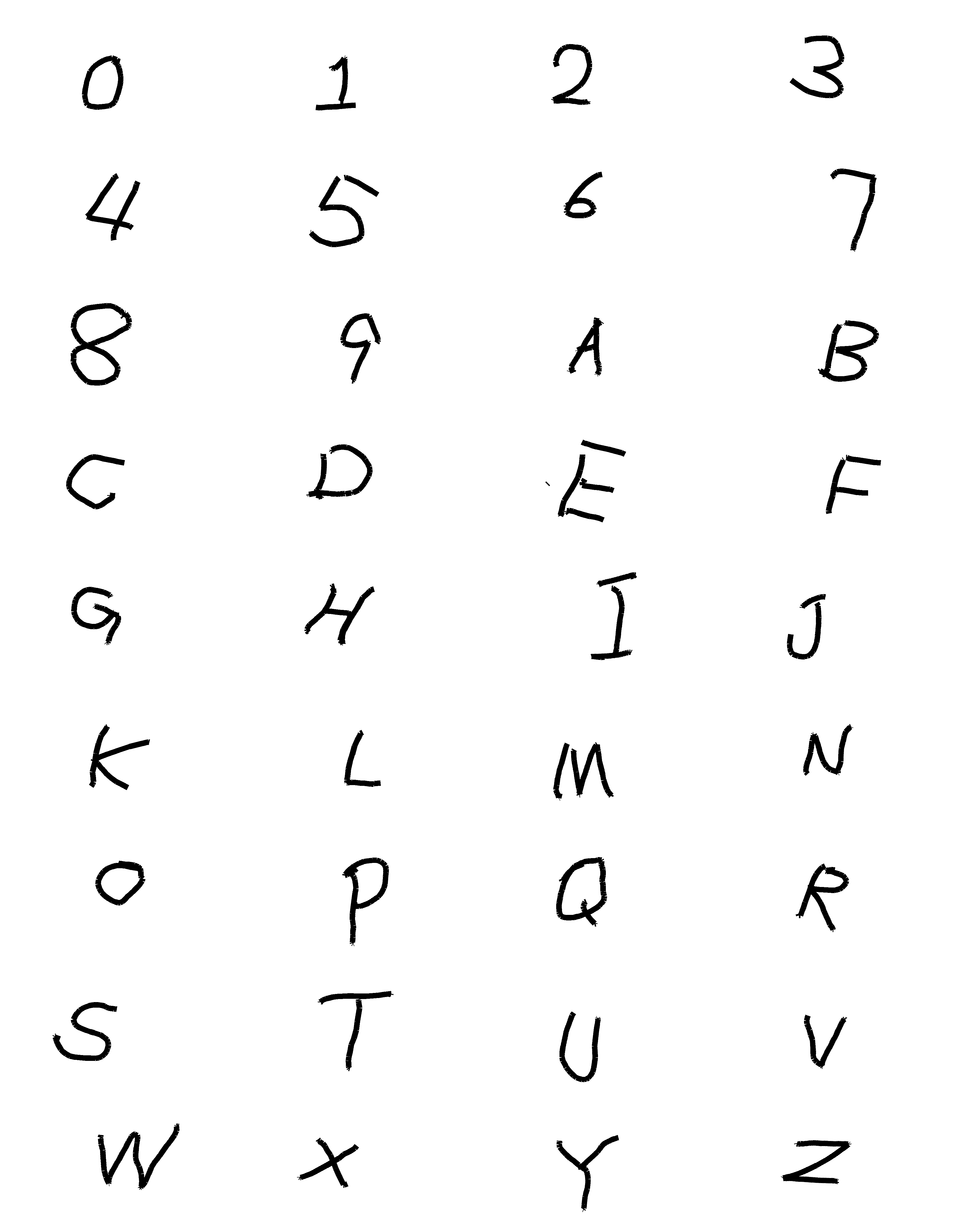}
    \caption{Example characters written by different users.}
    \label{fig:hand_subfig_b}
  \end{subfigure}

  \medskip
  \begin{subfigure}[t]{0.72\textwidth}
    \centering
    \includegraphics[width=\linewidth]{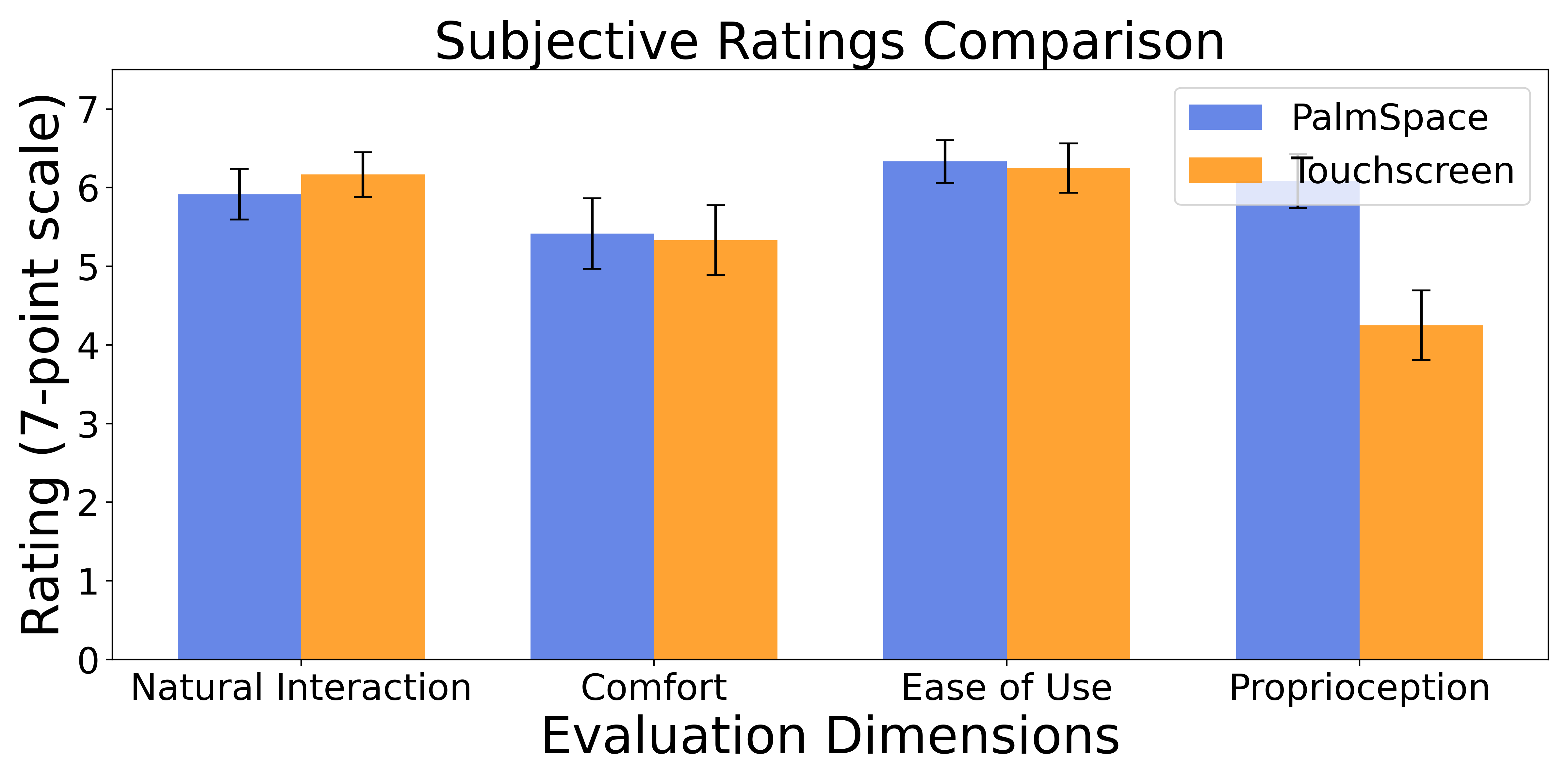}
    \caption{Subjective ratings for PalmSpace and touchscreen input.}
    \label{fig:subject_rating}
  \end{subfigure}

  \caption{Handwriting recognition and subjective evaluation results.}
  \label{fig:hands}
  \Description{The top row contains a 36-class handwritten-character confusion matrix and representative character traces from multiple participants. The bottom panel compares participant ratings of PalmSpace and touchscreen input.}
\end{figure}

In the handwriting experiment, PalmSpace achieved an average recognition accuracy of 92.5\%, with 26 of the 36 characters recognized at 100\%. The confusion matrix is shown in Fig. \ref{fig:hand_subfig_a}. However, some characters still exhibited significant confusion cases, particularly the digit `0,' which was often misidentified as the similarly shaped letters `D' or `O'. Successfully recognized example results are shown in Fig. \ref{fig:hand_subfig_b}, where most users adopted a landscape orientation of the palm as the writing surface. 

For comparison, touchscreen input achieved 94.58\% accuracy. Statistical analysis (Shapiro-Wilk and repeated-measures ANOVA) showed no significant difference between the two methods ($F_{1,11}=1.21, p=0.295$). PalmSpace’s slightly lower mean accuracy can be attributed to palm curvature, which makes it harder to maintain stable strokes than on a flat touchscreen.

Overall, PalmSpace demonstrated handwriting performance comparable to gold-standard devices (i.e., touchscreens), confirming its potential for precise, free-space handwriting input. In addition, as shown in Table \ref{tab:related_work}, it is calibration-free (users could use the system immediately without individual setup) and supports 2D finger tracking and touch detection for high-precision natural handwriting, unlike systems that constrain users to continuous stroke input \cite{he2025writingring}.

\subsection{Qualitative Feedback}
Participants consistently reported that PalmSpace offered stronger proprioceptive cues than the touchscreen during eyes-free use (Fig.~\ref{fig:subject_rating}). Several participants noted that localizing positions on the palm felt easier than localizing positions on a visually unavailable screen. They also mentioned that PalmSpace may be especially promising in relaxed or immersive scenarios such as large-screen viewing or mixed-reality interaction, where taking out a phone or shifting visual attention to a display would be undesirable.

\FloatBarrier

\section{Supplementary Details for Multi-Finger User Study}
\label{app:multifinger}

\subsection{Interaction Parameterization for Pinch}
\label{app:multifinger-parameterization}
For continuous manipulation in \textit{Pinch Mode}, PalmSpace maps the predicted interaction point $(x_{\mathrm{curr}}, y_{\mathrm{curr}})$ relative to the palm center $(C_x, C_y)$ into a polar representation. Specifically, we compute:
\begin{equation}
\begin{aligned}
d &= \sqrt{(x_{\mathrm{curr}} - C_x)^2 + (y_{\mathrm{curr}} - C_y)^2}, \\
\theta &= \operatorname{atan2}\!\left(-(y_{\mathrm{curr}} - C_y),\, x_{\mathrm{curr}} - C_x\right).
\end{aligned}
\end{equation}
where $d$ denotes the radial distance from the palm center and $\theta$ denotes the angular position. In our interaction design, $d$ is used to control scaling and $\theta$ is used to control rotation. This representation allows PalmSpace to decompose a single palm-based multi-finger gesture into two continuous control channels. 

\subsection{Transfer Functions for Continuous Control}
\label{app:multifinger-transfer}
To stabilize continuous interaction and improve controllability, we implemented separate transfer functions for scaling and rotation.

\subsubsection{Scaling Control}
The scaling factor is updated incrementally according to the change in radial distance: $\Delta d = d_{\mathrm{curr}} - d_{\mathrm{start}}$. We apply a sensitivity coefficient $\lambda_{\mathrm{zoom}} = 0.3$ to modulate the control-display ratio. The updated scale is computed as:
\begin{equation}
    S_{\mathrm{new}} = \operatorname{clip}(S_{\mathrm{base}} + \Delta d \cdot \lambda_{\mathrm{zoom}}, [0.2, 5.0])
\end{equation}
The clipping range $[0.2, 5.0]$ prevents extreme visual distortion. This incremental update rule improves stability compared with directly mapping absolute radial distance to scale.

\subsubsection{Rotation Control}
Rotation is driven by the angular displacement $\Delta \theta = \theta_{\mathrm{curr}} - \theta_{\mathrm{start}}$. Because $\operatorname{atan2}$ is discontinuous at $\pm 180^\circ$, we applied a wrapping compensation mechanism:
\begin{equation}
\Delta \theta \leftarrow
\begin{cases}
\Delta \theta - 360^\circ, & \text{if } \Delta \theta > 180^\circ \\
\Delta \theta + 360^\circ, & \text{if } \Delta \theta < -180^\circ \\
\Delta \theta, & \text{otherwise}
\end{cases}
\end{equation}
To support ergonomic operation with relatively small physical finger motion, we further apply a rotational gain factor $\gamma_{\mathrm{rot}} = 4.0$. The final rotation angle is updated as:
\begin{equation}
    \Phi_{\mathrm{new}} =
    \left(\Phi_{\mathrm{base}} + \Delta \theta \cdot \gamma_{\mathrm{rot}}\right)
    \pmod{360^\circ}
\end{equation}

\section{Supplementary Details for Real-World Outdoor Robustness}
\label{app:outdoor_study}

This section provides detailed methodologies, datasets, and quantitative ablation results for the outdoor evaluation scenarios discussed in the main paper.

\begin{figure*}[ht]
  \centering
  \includegraphics[width=0.9\linewidth]{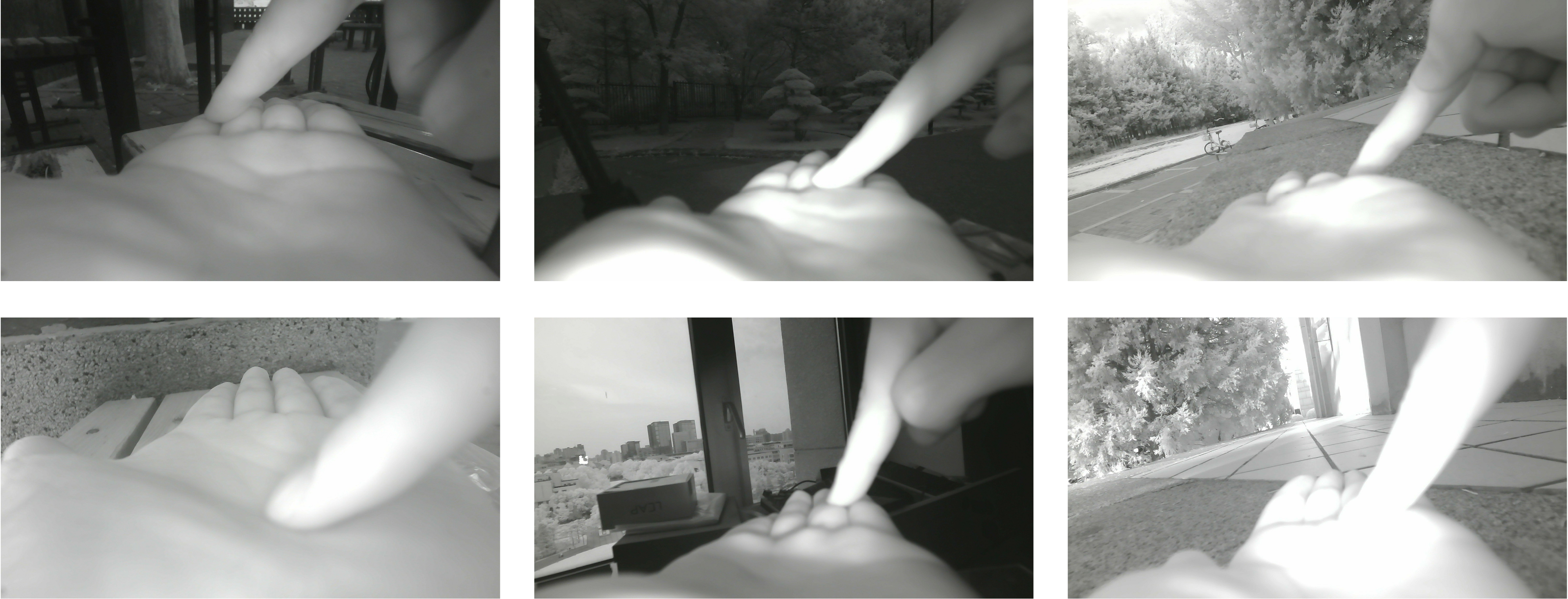}
  \caption{Sample images from various outdoor data collection scenarios.}
  \label{fig:sampleimage_app}
  \Description{A grid of infrared wrist-camera frames shows palm interactions collected outdoors across varied locations, daylight levels, and weather conditions.}
\end{figure*}

\subsection{Daytime Data Collection and Protocol}
The daytime experiment was designed to directly assess the model's generalization capabilities in natural outdoor settings. Data collection was performed at locations arranged with the users, encompassing a variety of environments such as sunny and overcast conditions, parks, streets, and balconies. In total, \textbf{14 sets} of in-the-wild data were collected, comprising \textbf{10,388 contact images} and \textbf{3,514 non-contact images}. During acquisition, participants were instructed to touch the entire capturable area of their palm without constraints on finger posture to simulate real-world usage. Fig. \ref{fig:sampleimage_app} illustrates sample images from these scenarios.

\subsection{Network Ablation Study}
We compare the unified model with architectural baselines and independently trained task-specific models. All zero-shot models are trained on the indoor dataset and directly evaluated on the outdoor dataset described above. For the fine-tuned condition, the outdoor data are partitioned using leave-one-participant-out cross-validation. The quantitative comparison is reported with the main ablation analysis in Table~\ref{tab:ablation}(b).

\subsection{Note on Nighttime Ground Truth}
In the nighttime scenario, standard RGB-based visual ground truth methods were found to be unreliable. Specifically, MediaPipe's hand tracking performance degrades significantly when ambient illuminance falls below 5 lux, leading to inaccurate keypoint detection. Therefore, the nighttime evaluation relied on the task completion time of the Fitts' Law study rather than direct geometric error measurement.

\section{Supplementary Details for SUS Feedback}
\label{app:multifinger-sus}
We assessed perceived usability using the System Usability Scale (SUS) \cite{brooke1996sus}. The overall SUS score was 72.5 $\pm$ 14.0 (median = 70.0, IQR = 16.0), corresponding to a \textit{Good} usability level \cite{bangor2008empirical}. Participants gave high ratings to statements regarding ease of use ($M = 5.00$) and learnability ($M = 5.17$), and agreement with negatively worded items remained low.

At the item level, participants gave high ratings to positively worded statements such as:
\begin{itemize}
    \item ``I thought the product was easy to use'' ($M = 5.00$),
    \item ``I imagine that most people would learn to use this product very quickly'' ($M = 5.17$),
    \item ``I felt very confident using the product'' ($M = 5.00$).
\end{itemize}
These responses indicate favorable perceptions of ease of use, learnability, and confidence during interaction.

Agreement with negatively worded items remained relatively low, including:
\begin{itemize}
    \item ``I found the product unnecessarily complex'' ($M = 2.09$),
    \item ``I found the product very awkward to use'' ($M = 2.33$),
    \item ``I needed to learn a lot of things before I could get going with this product'' ($M = 2.17$).
    \item ``I think that I would need the support of a technical person to be able to use this product'' ($M = 3.18$).

  \end{itemize}
These results suggest limited perceived friction and a relatively smooth learning curve.

\end{document}